\documentclass[a4paper]{article}

\usepackage{empheq}
\usepackage{float} 
\usepackage{microtype}
\usepackage{enumitem}
\usepackage{booktabs}
\usepackage{hyperref}

\newcommand{\bs}{\boldsymbol}

\title{\textbf{Thermomechanical Modeling of Microstructure Evolution Caused by Strain-Induced Crystallization}}
\author{\textbf{Serhat Aygün\thanks{Correspondence: serhat.ayguen@tu-dortmund.de; Tel.: +49-231-755-7902} \, and Sandra Klinge}  \\ \\
	{\normalsize Institute of Mechanics, TU Dortmund University,} \\
	{\normalsize Leonhard-Euler-Strasse 5, 44227 Dortmund, Germany}\\ \\
	}

\date{{\small \today}}

\begin{document}

\maketitle

\begin{abstract}
The present contribution deals with the thermomechanical modeling of the strain-induced crystallization in unfilled polymers. This phenomenon significantly influences mechanical and thermal properties of polymers and has to be taken into consideration when planning manufacturing processes as well as  applications of the final product. In order to simultaneously capture both kinds of effects, the model proposed starts by introducing a triple decomposition of the deformation gradient and furthermore uses thermodynamic framework for material modeling based on the Coleman--Noll procedure and minimum principle of the dissipation potential, which requires suitable assumptions for the Helmholtz free energy and the dissipation potential. The chosen setup yields  evolution equations which are able to simulate the  formation and the degradation of crystalline regions accompanied by the temperature change during a cyclic tensile test. The boundary value problem corresponding to the described process includes the balance of  linear momentum and balance of energy and serves as  a basis for the numerical implementation within an FEM code. The~paper closes with the numerical examples showing the microstructure evolution and temperature distribution for different material samples.
\end{abstract}

\paragraph{Keywords:}
thermomechanical modeling;   microstructure;  energy;  dissipation;  strain-induced crystallization; polymers; temperature; thermodynamic consistency

\section{Introduction}
The strain-induced crystallization (SIC) is a phenomenon of formation and degradation of tiny crystalline regions within the amorphous polymer structure due to the  strain variation. It has a significant influence on the production and application of the affected materials, which strongly motivates its  experimental investigation and numerical simulation. 
The experimental characterization of the process relies on techniques such as volume change measurements~\cite{10.5254/1.3525684}, electron microscopy~\cite{QU20095053}, small-angle X-ray scattering~\cite{Kojio2011}, and terahertz spectroscopy~\cite{sommer2016}. However, the~most frequently used technique is the in~situ wide-angle X-ray diffraction. Amongst others, the~method has been applied to study crystalline content and orientation~\cite{doi:10.5254/rct.13.86977}, the~SIC kinetics~\cite{TOSAKA2012864}, the~effects of dynamic loading~\cite{bruening2012}, and the distribution of the crystallite size~\cite{doi:10.1021/ma5006843}. Infrared thermography is also worth mentioning, particularly in the context of investigating thermal effects~\cite{SPRATTE201712,doi:10.1111/str.12256,10.1007/978-3-319-95074-7_11}.

\begin{figure}[H]
\centering
\includegraphics[width=\textwidth]{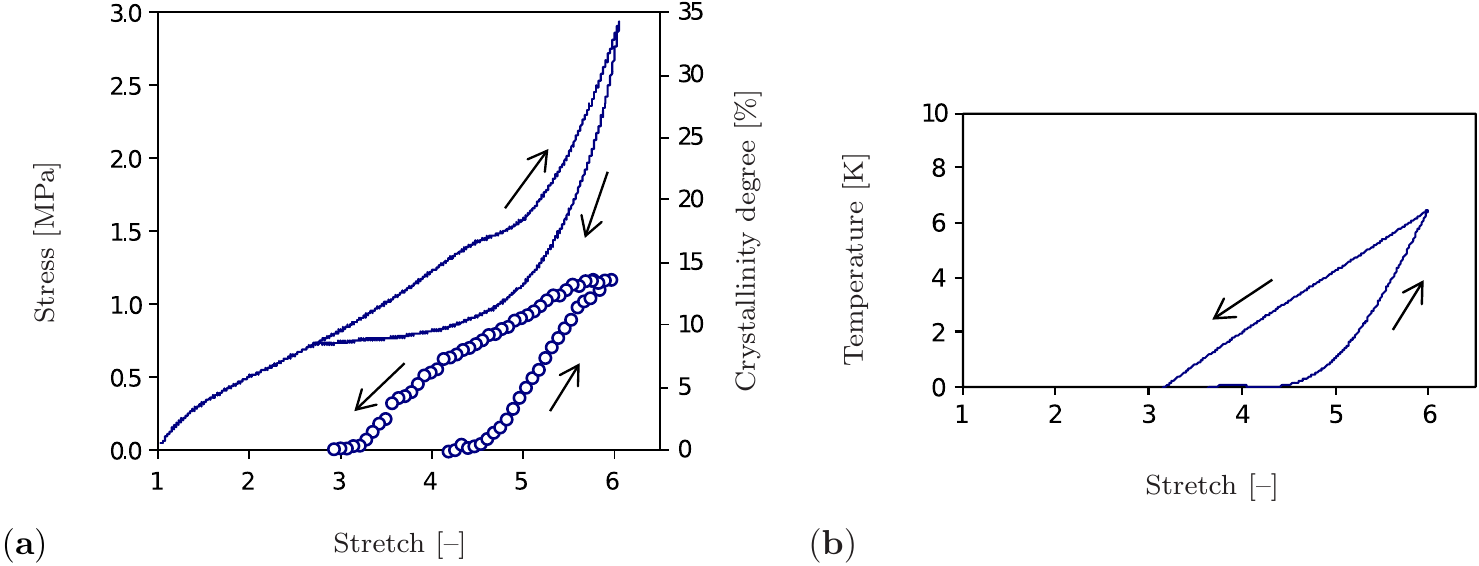}
  \caption{ {Uniaxial cyclic tensile test for a vulcanized unfilled natural rubber at room temperature and at a constant strain rate of $4.2 \times 10^{-3}$ s$^{-1}$. (\textbf{a}) Stress--stretch diagram (solid line) and crystallinity degree--stretch diagram (circle symbols). (\textbf{b}) Temperature--stretch diagram. Reprinted from~\cite{CANDAU2015244}, Copyright (2020), with permission from Elsevier.} }
  \label{fig1}
\end{figure}

The representative results of a cyclic tensile test performed for the unfilled natural rubber at constant ambient temperature and strain rate are shown in Figure~\ref{fig1}. Here, the~stress diagram (Figure~\ref{fig1}a) forms a hysteresis and thus shows the dissipative nature of the SIC phenomenon. The~volume fraction of the crystalline regions, the~so-called degree of crystallinity, is also used to show the crystallization process. According to Figure~\ref{fig1}a, crystalline regions begin to form after exceeding a stretch of $\lambda=4.3$. The~degree of crystallinity rises with increasing strain until saturation at about $\lambda=6.0$ is reached. A~further increase in total deformation may lead to inelastic deformations, which is not the subject of this study. The~decrease of the degree of crystallinity during the unloading phase is less intensive than its growth by loading. The~material becomes completely amorphous at a stretch of $\lambda=3.0$. It is also remarkable that the heat generated by the formation of crystalline regions leads to an increase in temperature, whereas the regression of the crystalline regions during the unloading phase causes a decrease in temperature. Accordingly, the~temperature diagram (Figure~\ref{fig1}b) has a form analogous to the one shown in Figure~\ref{fig1}a.

With regard to the modeling and simulation, the~first effort was made by Alfrey and Mark (1942)~\cite{alfrey1942} who investigated  the behavior of a single polymer chain. On~the basis of the this work, Flory (1947, 1949)~\cite{doi:10.1063/1.1746537,flory1949} formulated the classical thermodynamic theory of the SIC. In~these works, the~degree of crystallization is expressed by the change in crystallization temperature which, in~turn, depends on strain. At~a later stage, a~phenomenological expression to describe the growth of the crystalline phase was proposed by Doufas~et~al. (1999)~\cite{doufas1999} using a modified Avrami equation. The~work by Ahzi~et~al. (2003)~\cite{ahzi2003} extends the hyperelastic model of Boyce~et~al. (1993)~\cite{ARRUDA1993389} by the phenomenological expression of Doufas~et~al. (1999) in order to explain the growth of the crystalline phase. The~works by  Negahban (2000)~\cite{NEGAHBAN20002811} and Rao and Rajagopal (2001)~\cite{RAO20011149} used the large deformation theory to model the amorphous and crystalline phases separately. Afterwards, Tosaka~et~al. (2004)~\cite{tosaka2004} developed a micromechanical model in which shorter chains are completely stretched under tension and form nucleation sites for crystals. The~model by Kroon (2010)~\cite{kroon2010} deals with the anisotropic nucleation in unfilled rubbers. It  predicts both stress--stretch hysteresis and the development of the degree of crystallization, under~the assumption that the dissipative process is not  caused  by crystallization only but also by the viscoelastic behavior of the amorphous phase. An~advanced micromechanical continuum model for partially crystallized polymers was developed by Mistry and Govindjee (2014)~\cite{mistry2014}. In~their work,  the~micromechanical model is connected to the macroscopic scale by using the non-affine microsphere model.  The~model is able to quantitatively predict the macroscopic behavior of strain-crystallizing rubbers. Recently,  Nateghi~et~al. (2018)~\cite{nateghi2018} similarly proposed a micromechanical model that is incorporated into the affine microsphere model. In~a  contribution by Dargazany~et~al. (2014)~\cite{dargazany2014}, a~micromechanical model for SIC in filled rubbers  additionally considers inelastic properties of filled rubbers such as the Mullins effect, the~permanent setting effect, and the induced anisotropy. In~contrast to the aforementioned models that focus on the modeling of mechanical effects of the SIC, the~publication by Behnke~et~al. (2018)~\cite{BEHNKE201815} simulates the temperature and time dependency of SIC by taking the induced anisotropy into account. The~newer publications by Khi\^em and Itskov (2018)~\cite{KHIEM2018350} and Khi\^em~et~al. (2019)~\cite{khiem2019} focus on calorimetric effects, such as the temperature change due to SIC and the evolution of heat sources, whereas the work by Felder~et~al. (2019)~\cite{felder2019} proposes a thermomechanically coupled model related to the crystallization by cooling from the melt.

{
Molecular dynamics is an alternative approach to modeling the microstructural phenomena of SIC, as~presented in works by Nie (2015) \cite{Nie2015} and Yamamoto (2019) \cite{Yamamoto2019}. This  approach is suitable for investigations at  nanoscale since it directly simulates effects related to interatomic potentials or thermal fluctuations of atoms. However, the~approach is time-consuming and subject to a high computational effort. This strongly motivates the continuum mechanical modeling which intrinsically includes the aspects mentioned above and enables the efficient simulation of much larger samples than in the case of the molecular dynamics. Coupled with the multiscale strategies,  a~continuum mechanical model is even able to simulate practical applications.
}

As the previous overview shows, the~development of measurement techniques has already made a significant contribution to the study of the SIC process. Nevertheless, there are still open issues which have not yet been sufficiently clarified by experimental studies, since phenomena are related to the nanoscale and are thus not accessible by the experimental techniques. Amongst others, the~influence of the heat production on the shape, distribution, and mutual interaction of crystalline regions for high strain states can be pointed out. Moreover, already existing models mostly provide data on the effective material behavior without giving insight into the developed microstructure and the related temperature distribution. In~contrast to these strategies, the~present model treats the microstructural changes of the crystalline regions within the amorphous polymer matrix as well as the heat production of both phases. Our previous work~\cite{AYGUN2020129} presents a material model which depicts the mechanical characteristics of the SIC phenomenon in agreement with the experimental investigations { (Figure \ref{fig1})}. Based on this model, the~aim of the present contribution is to visualize the microstructural development and the change of temperature distribution within a material sample and to observe their interaction depending on external influences. The~approach proposed focuses on the unfilled polymers being the representatives of  nearly incompressible materials. 
%

The paper is structured as follows: Section~\ref{basconc} gives an overview of general concepts used for the thermomechanical modeling of the SIC. It introduces the finite deformation kinematics that rely on the multiplicative split of the deformation gradient into an elastic, crystalline, and thermal part and assumes the Arruda--Boyce model for elastically stored energy. The~same section  investigates the thermodynamic consistency, determines the energetically conjugated pairs, and uses the minimum principle of dissipation potential to derive evolution equations. Subsequently, it also defines the boundary value problem (BVP) including the balance of linear momentum and balance of energy. Whereas Section~\ref{basconc} deals with the general concepts, Section~\ref{thmemo} introduces specific assumptions related to the modeling of the SIC. Primarily, it defines the internal variables and proposes expressions for the Helmholtz  free energy and dissipation potential. Both are chosen such that the resulting evolution equations simulate the increase and the decrease of the crystalline regions and of the temperature during a cyclic test. Finally, Section~\ref{numimpl} provides details on the implementation of the material model into the FEM software FEAP, and~Section~\ref{results} demonstrates the application. Selected~numerical examples pertaining cyclic tensile loads visualize the microstructure evolution and the corresponding temperature distribution. At~first, an~academic example studies the material response of three crystalline regions with different initial regularity while growing and shrinking. Thereafter, the~model is applied to monitor the microstructure evolution and temperature distribution for a sample with a complex initial configuration. The~paper finishes with conclusions and an~outlook.

\section{Thermodynamically Consistent~Framework}
\label{basconc}
\vspace{-6pt}

\subsection{Kinematics within the Finite Thermomechanical~Framework}
\label{fin_kin}
\begin{figure}[H]
  \centering
  \includegraphics[width=0.7\textwidth]{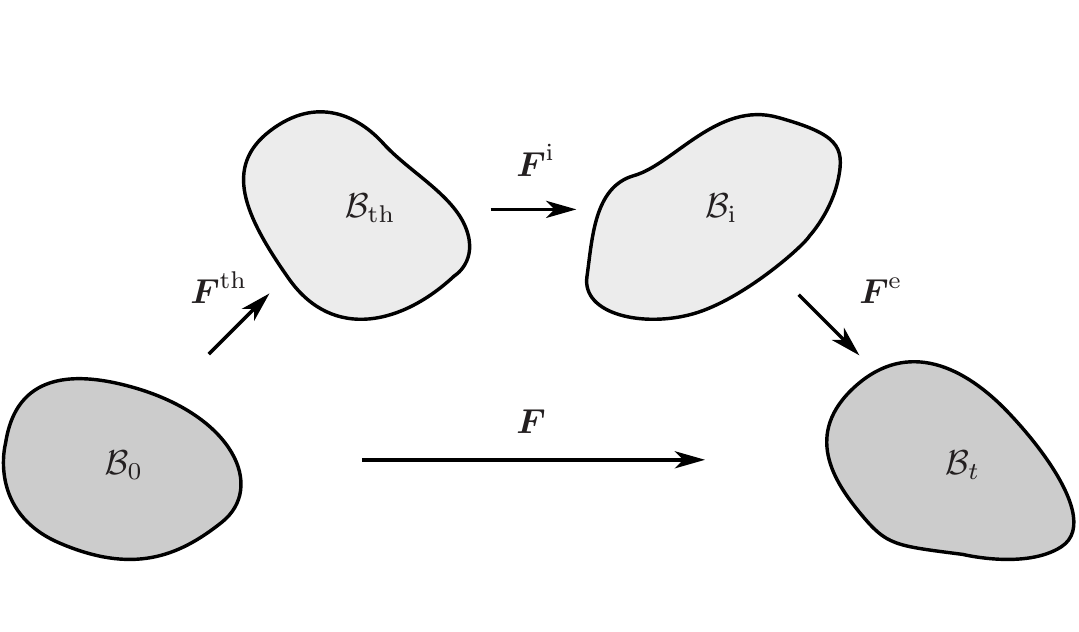}
  \caption{The multiplicative decomposition of the deformation gradient and the corresponding mapping between the initial, thermal intermediate, inelastic intermediate, and current~configuration.}
  \label{2iconf}
\end{figure}

The multiplicative  decomposition   of  the deformation gradient $\bs{F}$ into an elastic part $\bs{F}^{\mathrm{e}}$, an~inelastic part $\bs{F}^{\mathrm{i}}$, and a thermal part $\bs{F}^{\mathrm{th}}$
\begin{equation}
  \bs{F} = \bs{F}^{\mathrm{e}} \cdot \bs{F}^{\mathrm{i}} \cdot \bs{F}^{\mathrm{th}} 
  \label{defgrad}
\end{equation}
is an appropriate method  to simultaneously incorporate inelastic and thermal effects within the finite deformation theory~\cite{klinge2012, LIAO2020103263}. The~deformation gradient $\bs{F} $ is a two-point tensor that maps the initial configuration ($\mathcal{B}_0$) to the deformed configuration ($\mathcal{B}_t$). Formula \eqref{defgrad}, however, implies the existence of two additional intermediate configurations (Figure~\ref{2iconf}) such that $\bs{F}^{\mathrm{th}}$ maps the initial to the thermal intermediate configuration ($\mathcal{B}_{\mathrm{th}}$). The~latter is then related to the intermediate configuration $\mathcal{B}_{\mathrm{i}}$ by the inelastic deformation gradient $\bs{F}^{\mathrm{i}}$. Eventually, the~elastic deformation gradient $\bs{F}^{\mathrm{e}}$ maps the inelastic intermediate to the current configuration. It is also worth mentioning that the ordering of intermediate configurations is not uniquely prescribed. Hartmann (2012)~\cite{Hartmann2012}, for~example, defines the multiplicative decomposition into the thermal and mechanical deformation gradient in a reverse~order.

Based on assumption \eqref{defgrad}, the~elastic deformation gradient is written as
\begin{equation}
  \bs{F}^{\mathrm{e}} = \bs{F} \cdot {\bs{F}^{\mathrm{th}}}^{-1} \cdot {\bs{F}^{\mathrm{i}}}^{-1}
  \label{Fe}
\end{equation}
which furthermore yields the relationship for the corresponding rate
\begin{equation}
  \dot{\bs{F}}^{\mathrm{e}} = \dot{\bs{F}} \cdot {\bs{F}^{\mathrm{th}}}^{-1} \cdot {\bs{F}^{\mathrm{i}}}^{-1} + \bs{F} \cdot \dot{\overline{{{\bs{F}}^{\mathrm{th}}}^{-1}}} \cdot {\bs{F}^{\mathrm{i}}}^{-1} + \bs{F} \cdot {\bs{F}^{\mathrm{th}}}^{-1} \cdot \dot{\overline{{{\bs{F}}^{\mathrm{i}}}^{-1}}} \text{ .}
  \label{dFe}
\end{equation}

The relationship above depends on the time derivatives $\dot{\overline{{{\bs{F}}^{\mathrm{i}}}^{-1}}}$ and $\dot{\overline{{{\bs{F}}^{\mathrm{th}}}^{-1}}}$, which can be expressed in a more appropriate manner as will be shown by the example of $\dot{\overline{{{\bs{F}}^{\mathrm{i}}}^{-1}}}$.
The procedure starts by taking the time derivative of the identity tensor $\bs{I}$, which is expressed as a function of $\bs{F}^{\mathrm{i}}$
\begin{equation}
  \bs{F}^{\mathrm{i}} \cdot {\bs{F}^{\mathrm{i}}}^{-1} = \bs{I} \quad \Rightarrow \quad \dot{\overline{\left( \bs{F}^{\mathrm{i}} \cdot {\bs{F}^{\mathrm{i}}}^{-1} \right)}} = \dot{\bs{F}}^{\mathrm{i}} \cdot {\bs{F}^{\mathrm{i}}}^{-1} + \bs{F}^{\mathrm{i}} \cdot \dot{\overline{{{\bs{F}}^{\mathrm{i}}}^{-1}}} = \bs{0}\,.
  \label{dI_}
\end{equation}

Rearranging Equation \eqref{dI_}
 then yields
\begin{equation}
  \dot{\overline{{{\bs{F}}^{\mathrm{i}}}^{-1}}} = - {\bs{F}^{\mathrm{i}}}^{-1} \cdot \dot{\bs{F}}^{\mathrm{i}} \cdot {\bs{F}^{\mathrm{i}}}^{-1} \text{ .}
  \label{dFii}
\end{equation}

A similar procedure provides an analogous relationship for $\dot{\overline{{{\bs{F}}^{\mathrm{th}}}^{-1}}}$:
\begin{equation}
  \dot{\overline{{{\bs{F}}^{\mathrm{th}}}^{-1}}} = - {\bs{F}^{\mathrm{th}}}^{-1} \cdot \dot{\bs{F}}^{\mathrm{th}} \cdot {\bs{F}^{\mathrm{rh}}}^{-1} \text{ ,}
  \label{dFthi}
\end{equation}
such that the insertion of  Equations~\eqref{Fe}, \eqref{dFii}, and \eqref{dFthi} into \eqref{dFe} ends up in
\begin{equation}
  \dot{\bs{F}}^{\mathrm{e}} = \dot{\bs{F}} \cdot {\bs{F}^{\mathrm{th}}}^{-1} \cdot {\bs{F}^{\mathrm{i}}}^{-1} - \bs{F} \cdot {\bs{F}^{\mathrm{th}}}^{-1} \cdot \dot{\bs{F}}^{\mathrm{th}} \cdot {\bs{F}^{\mathrm{th}}}^{-1} \cdot {\bs{F}^{\mathrm{i}}}^{-1} - \bs{F}^{\mathrm{e}} \cdot \dot{\bs{F}}^{\mathrm{i}} \cdot {\bs{F}^{\mathrm{i}}}^{-1} \text{ ,}
\label{dotFe}
\end{equation}
 which is the final expression for the rate elastic deformations only depending on rates and on inverse values of basic quantities ${\bs{F}}$, ${\bs{F}^{\mathrm{i}}}$, and ${\bs{F}^{\mathrm{th}}}$.

{ The nearly incompressible elasticity of polymeric materials is considered in works by Flory (1961)~\cite{TF9615700829} and Doll~et~al. (2000) \cite{doll2000}. These authors introduce the decomposition of the deformation gradient $\bs{F}^{\mathrm{e}}$ into a volmetric part $\hat{\bs{F}}^{\mathrm{e}}$ and a deviatoric part $\bar{\bs{F}}^{\mathrm{e}}$, both depending on the determinant $J^{\mathrm{e}}$:
\begin{equation}
  \bs{F}^{\mathrm{e}} = \hat{\bs{F}}^{\mathrm{e}} \cdot \bar{\bs{F}}^{\mathrm{e}} \text{ ,}
	\qquad\quad
	\hat{\bs{F}}^{\mathrm{e}} := J^{\mathrm{e}^{\frac{1}{3}}} \, \bs{I} \text{ ,} 
	\qquad \quad
	\bar{\bs{F}}^{\mathrm{e}} := J^{\mathrm{e}^{-\frac{1}{3}}} \, \bs{F}^{\mathrm{e}} \text{ ,}  
	\qquad\quad
	J^{\mathrm{e}} = \mathrm{det}(\bs{F}^{\mathrm{e}}) \text{ .}
  \label{voldev}
\end{equation}

 For the isochoric deformations, it holds that $J^{\mathrm{e}}= 1$ since the determinant $J^{\mathrm{e}}$ represents the measure of the elastic volumetric changes. 
}
\subsection{The Arruda--Boyce Elastic~Energy}
\label{AB_energy}
{ Nowadays, there are plenty of different models for the elastic energy part, as~is shown in the overview by Steinmann~et~al. (2012) \cite{Steinmann2012}. However, the~present contribution assumes the Arruda--Boyce model relying on the statistical mechanics of idealized polymer chains. This  micromechanical model is suitable for simulating the large elastic deformations and excellently fits to the chosen thermodynamic framework and numerical concept. Moreover, it only depends on  two material parameters which are experimentally easily accessible and which are already known for many standard polymers such as natural rubber~\cite{hoss2010} and unfilled silicone rubber~\cite{MANSOURI20144316}. Since it is well established and explored,  the~Arruda--Boyce model represents an ideal platform for the present analysis where the focus is set on the investigation of dissipative aspects of the process rather than on the elastic effects. 

The Arruda--Boyce energy $\Psi^{\mathrm{AB}}$ originally has the form
\begin{equation}
  \Psi^{\mathrm{AB}} (\bar{I}_1) = \mu \, \lambda_m^2 \left( \frac{\lambda_{\mathrm{chain}}}{\lambda_m} \beta + \mathrm{ln} \frac{\beta}{\mathrm{sin}(\beta)} \right) \text{ ,} \quad \beta = L^{-1}\left( \frac{\lambda_{\mathrm{chain}}}{\lambda_m}  \right) \text{ ,} \quad \lambda_{\mathrm{chain}} = \sqrt{\frac{\bar{I}_1}{3}} \text{ ,}
  \label{ab_orig}
\end{equation}
where $\bar{I}_1$ is the  first invariant of the deviatoric elastic right Cauchy--Green tensor $\bar{\bs{C}}^{\mathrm{e}}$
\begin{equation} 
  \bar{I}_1 = \mathrm{tr}(\bar{\bs{C}}^{\mathrm{e}}) \text{ ,} 
	\qquad\quad \bar{\bs{C}}^{\mathrm{e}} = ({\bar{\bs{F}}^{\mathrm{e}}})^T \cdot \bar{\bs{F}}^{\mathrm{e}} = J^{\mathrm{e}^{-\frac{2}{3}}} \, ({\bs{F}^{\mathrm{e}}})^T \cdot \bs{F}^{\mathrm{e}} \text{ .} 
\end{equation} 

In Equation \eqref{ab_orig}, $\mu$ denotes the shear modulus, $\lambda_m$ is the limiting network stretch, $\lambda_{\mathrm{chain}}$ is the chain stretch depending on the deviatoric first invariant, and $\beta$ denotes the inverse Langevin function which is related to the energy of a single random chain.} The latter function cannot be expressed explicitly and is usually approximated by the Taylor series truncated up to the certain order~\cite{TF9545000881,doi:10.1177/1081286511429886}. The~present work assumes that an approximation including three terms of the Taylor series provides sufficient accuracy. In~this case, the~Arruda--Boyce energy takes the~form

{
\begin{equation}
  \Psi^{\mathrm{AB}} (\bar{I}_1) = \frac{\bar{\mu}}{2} \left[ \left( \bar{I}_1 - 3 \right) + \frac{1}{10 \, \lambda_m^2} \left( \bar{I}_1^2 - 9 \right) + \frac{11}{525 \, \lambda_m^4} \left( \bar{I}_1^3 - 27 \right) \right] \text{,}
    \label{ab}
\end{equation}
where parameter $\bar{\mu}$ is determined from the consistency condition of model \eqref{ab} with the linear elasticity theory for small strains. According to~\cite{horgan2010,doi:10.1063/1.5063384}, this condition is expressed as
\begin{equation}
  \left. \frac{\partial \Psi^{\mathrm{AB}}}{\partial \bar{I}_1} \right\vert_{\bar{I}_1 = 3} = \frac{\mu}{2} \text{ ,}
  \label{mu1}
\end{equation}
and requires the evaluation of the derivative of Equation \eqref{ab} at $\bar{I}_1 = 3$
\begin{equation}
  \left. \frac{\partial \Psi^{\mathrm{AB}}}{\partial \bar{I}_1} \right\vert_{\bar{I}_1 = 3} = \frac{\bar{\mu}}{2} \left( 1 + \frac{3}{5 \, \lambda_m^2} + \frac{99}{175 \, \lambda_m^4} \right) \text{ .}
  \label{mu2}
\end{equation}

Finally, the~implementation of Equation   \eqref{mu2}  in Equation \eqref{mu1} yields the expression for $\bar{\mu}$
\begin{equation}
  \bar{\mu} = \mu \left(1 + \frac{3}{5 \, \lambda_m^2} + \frac{99}{175 \, \lambda_m^4}\right)^{-1} \text{ .}
\end{equation} }
\vspace{-6pt}

Alternatively to the Taylor series, a~range of Pad{\'e} approximations can be applied  for the numerical evaluation of the inverse Langevin function. These approximations have different degrees of accuracy and complexity as discussed in the review papers by Jedynak~\cite{Jedynak} and Carroll~\cite{doi:10.1098/rsta.2018.0067}.

The original form of the Arruda--Boyce model corresponds to  the incompressible material behavior; however, it can be easily extended to capture the influence of compressibility by adding a volumetric part $\Psi^{\mathrm{vol}}$. In~that case, the~total elastic energy $\Psi^{\mathrm{e}}$ turns into
{
\begin{equation}
  \Psi^{\mathrm{e}} (\bar{I}_1, J^{\mathrm{e}}) = \Psi^{\mathrm{AB}} (\bar{I}_1) + \Psi^{\mathrm{vol}} (J^{\mathrm{e}}) \text{ .}
  \label{psivoldev}
\end{equation}

The chosen expression for the volumetric part of the energy is proposed by Simo and Taylor (1991)~\cite{SIMO1991273} and represents a special case of the Ogden model~\cite{doi:10.1098/rspa.1972.0096}
\begin{align}
  &\Psi^{\mathrm{vol}} (J^{\mathrm{e}}) = \frac{K}{4} \left( (J^{\mathrm{e}})^2 - 1 - 2 \, \mathrm{ln}(J^{\mathrm{e}}) \right) \text{ ,} 
\label{simoa} \\
  &\frac{\partial \Psi^{\mathrm{vol}} }{\partial J^{\mathrm{e}}} = \frac{K}{2} \left( J^{\mathrm{e}} - \frac{1}{J^{\mathrm{e}}} \right) \text{ ,}
  \label{simob}
\end{align}
where $K$ denotes the bulk modulus. Assumption \eqref{simoa} is  physically motivated with regard to the energy- and stress-free state in the reference configuration since it holds
\begin{equation}
  \Psi^{\mathrm{vol}} (J^{\mathrm{e}} = 1) = 0 \text{ ,} 
	\qquad\quad
	\frac{\partial \Psi^{\mathrm{vol}} }{\partial J^{\mathrm{e}}} (J^{\mathrm{e}} = 1) = 0 \text{ .}
  \label{psivol}
\end{equation} 

Moreover, the~energy function \eqref{simoa} is convex and coercive in $J^{\mathrm{e}}$. The~latter implies that the energy tends to infinity $\Psi^{\mathrm{vol}} \rightarrow \infty$ for high compression and tension modes, i.e.,~for the cases where $J^{\mathrm{e}} \rightarrow 0$ and $J^{\mathrm{e}} \rightarrow \infty$. A~comprehensive comparison of various models for the volumetric energy with respect to their mathematical and physical properties is shown by Hartmann and Neff (2003) \cite{HARTMANN20032767}, whereas the implementation of these approaches for nearly incompressible materials is discussed by Kadapa and Hossain (2020) \cite{doi:10.1080/15376494.2020.1762952}.

}

\subsection{Thermodynamic~Consistency}
\label{dissineq}
The second thermodynamic law in the form of the Clausius--Duhem inequality~\cite{kurth2002continuum} is chosen to investigate the thermodynamic consistency of the model:
\begin{equation}
  \mathcal{D} = - \dot{e} + \Theta \, \dot{\eta} + \frac{1}{\rho_0} \bs{P} \colon \dot{\bs{F}} + \mathcal{D}^{\mathrm{cond}} \ge 0 \text{,} \qquad \mathcal{D}^{\mathrm{cond}} = - \frac{1}{\rho_0 \, \Theta} \bs{q}_0 \cdot \nabla_{\bs{X}} \Theta \text{ .}
\label{di1}
\end{equation}

Here, $\dot{e}$ is the rate of internal energy, $\dot{\eta}$ is the rate of entropy, $\Theta$ is the temperature, $\rho_0$ denotes the density, $\bs{P} \colon \dot{\bs{F}}$ represents the internal power, $\bs{P}$ is the first Piola-Kirchhoff stress tensor, $\bs{X}$ are the coordinates in reference configuration, $\mathcal{D}^{\mathrm{cond}}$ is the heat conduction contribution to the dissipation $\mathcal{D}$, $\bs{q}_0$ is the heat flux vector, and index ``0'' refers to the initial configuration. Equation~\eqref{di1}
 has the standard solution determining the heat transfer through the material. This solution is known as the Fourier law and has the following form in the reference configuration~\cite{MAHNKEN20132003}
\begin{equation}
  \bs{q}_0 = - \lambda_{\theta} \, \mathrm{det}(\bs{F}) \, \bs{C}^{-1} \cdot \nabla_{\bs{X}} \Theta \mathrm{ ,} \quad \bs{C} = \bs{F}^T \cdot \bs{F} \text{,}
  \label{q0}
\end{equation}
where  $\lambda_{\theta}$ is  the thermal conductivity coefficient and $\bs{C}$ is the right Cauchy--Green deformation tensor.

An alternative formulation of the dissipation inequality \eqref{di1} is obtained by using the Legendre transformation of the internal energy yielding the Helmholtz free energy $\Psi $
\begin{equation}
  \Psi = e - \Theta \, \eta \text{ .}
  \label{legendre}
\end{equation}

Consequently,  the~internal energy rate can be expressed as a function of elastic energy rate $\dot{\Psi}$
\begin{equation}
  \dot{e} = \dot{\Psi} + \dot{\Theta} \, \eta + \Theta \, \dot{\eta} \text{ .}
  \label{dote}
\end{equation} 

The particular case studied in this contribution assumes the Helmholtz energy $\Psi$ as a function of three arguments: the elastic deformation gradient $\bs{F}^{\mathrm{e}}$, temperature $\Theta$, and an additional set of internal variables describing microstructural phenomena $\bs{\gamma}$, such that its rate turns into
\begin{equation}
  \dot{\Psi} = \frac{\partial \Psi}{\partial \bs{F}^{\mathrm{e}}} \colon \dot{\bs{F}}^{\mathrm{e}} + \frac{\partial \Psi}{\partial \Theta} \dot{\Theta} + \frac{\partial \Psi}{\partial \bs{\gamma}} \colon \dot{\bs{\gamma}} \text{ .}
  \label{dotpsi}
\end{equation}

Finally, the~insertion of Equations~\eqref{dotFe}, \eqref{dote}, and \eqref{dotpsi} into~\eqref{di1}
 yields an alternative form of the dissipation~inequality

\begin{equation}
  \begin{split}
  \mathcal{D} & = \left( \frac{1}{\rho_0} \bs{P} - \frac{\partial \Psi}{\partial \bs{F}^{\mathrm{e}}} \cdot \left({\bs{F}^{\mathrm{i}}}^{-1}\right)^T \cdot \left({\bs{F}^{\mathrm{th}}}^{-1}\right)^T \right) \colon \dot{\bs{F}} - \left( \frac{\partial \Psi}{\partial \Theta} + \eta \right) \dot{\Theta} \\
  & + \left( \left({\bs{F}^{\mathrm{th}}}^{-1} \right)^T \cdot \bs{F}^T \cdot \frac{\partial \Psi}{\partial \bs{F}^{\mathrm{e}}} \cdot \left( {\bs{F}^{\mathrm{i}}}^{-1} \right)^T \right) \colon \left( \dot{\bs{F}}^{\mathrm{th}} \cdot {\bs{F}^{\mathrm{th}}}^{-1} \right) \\
  & + \left( {\bs{F}^{\mathrm{e}}}^T \cdot \frac{\partial \Psi}{\partial \bs{F}^{\mathrm{e}}} \right) \colon \left( \dot{\bs{F}}^{\mathrm{i}} \cdot {\bs{F}^{\mathrm{i}}}^{-1} \right) - \frac{\partial \Psi}{\partial \bs{\gamma}} \colon \dot{\bs{\gamma}} + \mathcal{D}^{\mathrm{cond}} \ge 0 \text{ ,}
  \end{split}
  \label{di2}
\end{equation}
which is suitable for the application of the Coleman--Noll procedure and for determining the constitutive laws~\cite{Coleman1963}. Following the procedure mentioned, the~first two terms in Equation~\eqref{di2} yield definitions for the first Piola--Kirchhoff stress tensor and entropy
\begin{align}
  & \bs{P} = \rho_0 \frac{\partial \Psi}{\partial \bs{F}^{\mathrm{e}}} \cdot \left({\bs{F}^{\mathrm{i}}}^{-1}\right)^T \cdot \left({\bs{F}^{\mathrm{th}}}^{-1}\right)^T  \text{,} 
  \label{pi_etaa}\\ 
  & \eta = - \frac{\partial \Psi}{\partial \Theta} \text{ ,}
  \label{pi_etab}
\end{align}
whereas the remaining terms enable the definition of thermodynamically conjugated pairs. The~velocity gradient and Mandel stress tensor related to the thermal intermediate configuration are the first conjugated pair
\begin{align}
 &\bs{L}^{\mathrm{th}} := \dot{\bs{F}}^{\mathrm{th}} \cdot {\bs{F}^{\mathrm{th}}}^{-1} \text{ ,}
 \label{Dtha}\\ 
 &\bs{M}^{\mathrm{th}} := \left({\bs{F}^{\mathrm{th}}}^{-1} \right)^T \cdot \bs{F}^T \cdot \frac{\partial \Psi}{\partial \bs{F}^{\mathrm{e}}} \cdot \left( {\bs{F}^{\mathrm{i}}}^{-1} \right)^T
  \label{Dthb}
\end{align}
and velocity gradient and Mandel stress tensor related to the inelastic intermediate configuration represent the second conjugated pair
\begin{align}
 &\bs{L}^{\mathrm{i}} := \dot{\bs{F}}^{\mathrm{i}} \cdot {\bs{F}^{\mathrm{i}}}^{-1} \text{ ,} 
\label{Dia}\\
 &\bs{M}^{\mathrm{i}} := {\bs{F}^{\mathrm{e}}}^T \cdot \frac{\partial \Psi}{\partial \bs{F}^{\mathrm{e}}} \text{ .}
  \label{Dib}
\end{align}

In addition, both Mandel stress tensors (Equations~\eqref{Dthb} and \eqref{Dib}) can be related to each other by considering the multiplicative decomposition of the deformation gradient (Equation~\eqref{defgrad})
\begin{equation}
  \bs{M}^{\mathrm{th}} = \left({\bs{F}^{\mathrm{th}}}^{-1} \right)^T \cdot {\bs{F}^{\mathrm{th}}}^T \cdot {\bs{F}^{\mathrm{i}}}^T \cdot {\bs{F}^{\mathrm{e}}}^T \cdot \frac{\partial \Psi}{\partial \bs{F}^{\mathrm{e}}} \cdot \left( {\bs{F}^{\mathrm{i}}}^{-1} \right)^T =  {\bs{F}^{\mathrm{i}}}^T \cdot \bs{M}^{\mathrm{i}} \cdot \left( {\bs{F}^{\mathrm{i}}}^{-1} \right)^T \text{ .}
	  \label{Dth2}
\end{equation}

According to the previous relationship, the~Mandel stress tensor is mapped  from the intermediate inelastic configuration to the intermediate thermal configuration, which corresponds to a  pullback~operation.

Eventually, the~last term in inequality \eqref{di2} provides the definition for the driving force corresponding to the rate of internal variables  $\dot{\bs{\gamma}}$
\begin{equation}
  \bs{q}_{\gamma} := - \frac{\partial \Psi}{\partial \bs{\gamma}} \text{ .}
  \label{qgam}
\end{equation}

Using the new notation (Equations~\eqref{Dtha}--\eqref{Dib} and  \eqref{qgam}), the dissipation inequality \eqref{di2} is written in a shorter form
\begin{equation}
  \mathcal{D} = \mathcal{D}^{\mathrm{th}} + \mathcal{D}^{\mathrm{i}} + \mathcal{D}^{\gamma} + \mathcal{D}^{\mathrm{cond}} \ge 0 \text{ ,}
  \label{di3a}
\end{equation}
which includes the contribution due to thermal deformations $\mathcal{D}^{\mathrm{th}}$,  the~contribution due to the inelastic deformations $\mathcal{D}^{\mathrm{i}}$,  the~contribution due to microstructural changes $\mathcal{D}^{\gamma}$, and the contribution due to the heat conduction $\mathcal{D}^{\mathrm{cond}}$. The~single contributions are defined as follows:
\begin{equation}
  \mathcal{D}^{\mathrm{th}} = \bs{M}^{\mathrm{th}} \colon \bs{L}^{\mathrm{th}} \ge 0 \text{ ,} \quad \mathcal{D}^{\mathrm{i}} = \bs{M}^{\mathrm{i}} \colon \bs{L}^{\mathrm{i}} \ge 0 \text{ ,} \quad \mathcal{D}^{\gamma} = \bs{q}_{\gamma} \colon \dot{\bs{\gamma}} \ge 0 \text{ ,}
  \label{di3b}
\end{equation}
where the non-negativity of each term is required~separately. 

\subsection{Derivation of Evolution~Equations}
\label{mindiss}
Evolution equations for internal variables are an essential part of material models for inelastic processes. Their derivation is a challenging task such that different approaches have been established for this purpose~\cite{hackl2008}. The~present contribution follows the  minimum principle of dissipation potential~\cite{doi:10.1098/rspa.2014.0994} which is expressed as follows
\begin{equation}
  \min \{ \mathcal{L} = \dot{\Psi} + \Delta \, | \, \dot{\bs{\gamma}} \} \text{ .}
  \label{min}
\end{equation}

According to this principle, the~minimization of the Lagrangian $\mathcal{L}$ including the Helmholtz energy rate $\dot{\Psi}$ (Equation~\eqref{dotpsi}) and the dissipation potential $\Delta$ leads to the evolution laws for the internal variables ${\bs{\gamma}}$. Bearing in mind that the Helmholtz energy $\Psi$ is a function of three arguments $\bs{F}^{\mathrm{e}}$, $\Theta$, and  $\bs{\gamma}$, the~Lagrange function can be written in an extended form
\begin{equation}
 \mathcal{L} = \frac{\partial \Psi}{\partial \bs{F}^{\mathrm{e}}} \colon \dot{\bs{F}}^{\mathrm{e}} + \frac{\partial \Psi}{\partial \Theta} \dot{\Theta} + \frac{\partial \Psi}{\partial \bs{\gamma}} \colon \dot{\bs{\gamma}} + \Delta(\dot{\bs{\gamma}}) \,.
 \label{min2}
\end{equation}

Now, the~minimization of Equation~\eqref{min2} with respect to $\dot{\bs{\gamma}}$ reads the conditions
\begin{equation}
  \frac{\partial \mathcal{L}}{\partial \dot{\bs{\gamma}}} = \frac{\partial \Psi}{\partial \bs{\gamma}} + \frac{\partial \Delta}{\partial \dot{\bs{\gamma}}} = \bs{0} \quad \Rightarrow \quad \frac{\partial \Delta}{\partial \dot{\bs{\gamma}}} = - \frac{\partial \Psi}{\partial \bs{\gamma}} \text{ ,}
  \label{min3}
\end{equation}
and a comparison of Equations~\eqref{qgam} and \eqref{min3}
 leads to the conclusion that the driving force of an internal variable is equal to the derivative of the dissipation potential with respect to the same quantity
\begin{equation}
  \bs{q}_{\bs{\gamma}} = \frac{\partial \Delta}{\partial \dot{\bs{\gamma}}} \text{ .}
  \label{qgam2}
\end{equation}

Equations~\eqref{min}--\eqref{qgam2} define a generic procedure which will be used to derive equations driving the microstructural changes associated with the SIC (Section \ref{assumdiss}).

\subsection{Balance~Equations}
Different from a BVP, which is related to a purely mechanical problem, the~BVP corresponding to a thermomechanical problem includes two differential equations, namely the balance of linear momentum and the balance of  energy. These differential equations are accompanied by the suitable boundary conditions as follows~\cite{reese_govindjee,REESE2003909}:
\begin{eqnarray}
 \mathrm{Div} \bs{P} + \rho_0 \, \bs{b} = \bs{0} && \text{in } \,  \mathcal{B} \text{ ,}\label{balance1}\\
\dot{e} + \frac{1}{\rho_0} \mathrm{Div}(\bs{q}_0) = \frac{1}{\rho_0} \bs{P} \colon \dot{\bs{F}} + \, r_{\Theta}  && \text{in } \, \mathcal{B} \text{ ,} \label{balance2}\\
 \bs{P} \cdot \bs{n}=\bar{\bs{t}} && \text{on } \, \partial \mathcal{B}^{t} \text{ ,} \\
\bs{u} = \bar{\bs{u}} && \text{on } \, \partial \mathcal{B}^{u} \text{,} \\
-\bs{q}_0 \cdot \bs{n}= \bar{q}_{\Theta} && \text{on } \, \partial \mathcal{B}^{q} \text{ ,} \\
 \Theta = \bar{\Theta} && \text{on } \, \partial \mathcal{B}^{\Theta} \text{ .}\label{balance_end}
\end{eqnarray}

Here, $\bs{b}$ is the body force, $r_{\Theta}$ is the heat source, $\bs{q}_0$ is the heat flux vector defined by the Fourier law \eqref{q0}, and $\bs{n}$ is the surface normal vector. The~balance of linear momentum \eqref{balance1} is supplemented by the Neumann and Dirichlet boundary conditions expressed in terms of prescribed tractions $\bar{\bs{t}}$, displacements $\bs{u}$ and prescribed displacements $\bar{\bs{u}}$. The~corresponding boundary parts are denoted by $\mathcal{B}^{t}$ and $\mathcal{B}^{u}$, respectively. Similarly, the~balance of energy \eqref{balance2} is accompanied by boundary conditions relating the heat flux $\bs{q}_0$ and temperature $\Theta$ to the prescribed values $\bar{q}_{\Theta}$ and $\bar{\Theta}$ acting on the boundary parts $\mathcal{B}^{q}$ and $\mathcal{B}^{\Theta}$.

Internal energy rate $\dot{e}$ in Equation~\eqref{balance2} is defined by Equation~\eqref{dote} and thus requires a more precise study of entropy rate $\dot{\eta}$. To~this end, definition~\eqref{pi_etab}
 is used along  with the fact that the free energy is a function of three arguments:
\begin{equation}
\dot{\eta}=-\dot{\overline{\left(\frac{\partial\Psi}{\partial\Theta}\right)}} = - \left( \frac{\partial^2 \Psi}{\partial \Theta \, \partial \bs{F}^{\mathrm{e}}} \colon \dot{\bs{F}}^{\mathrm{e}} + \frac{\partial^2 \Psi}{\partial \Theta^2} \dot{\Theta} + \frac{\partial^2 \Psi}{\partial \Theta \, \partial \bs{\gamma}} \colon \dot{\bs{\gamma}} \right) \text{ .}
\label{doteta}
\end{equation}

Now, taking into account the definitions introduced in  Section~\ref{dissineq}, the~internal energy rate turns~into
\begin{equation}
  \dot{e} = \frac{1}{\rho_0} \bs{P} \colon \dot{\bs{F}} - \bs{M}^{\mathrm{th}} \colon \bs{L}^{\mathrm{th}} - \bs{M}^{\mathrm{i}} \colon \bs{L}^{\mathrm{i}} - \bs{q}_{\gamma} \colon \dot{\bs{\gamma}} + \Theta \, \dot{\eta} \text{ }
  \label{dote2}
\end{equation}
such that the final form of the energy balance  is obtained by inserting Equations~\eqref{doteta} and \eqref{dote2} into \eqref{balance2}
\begin{align}
& c_d \, \dot{\Theta} + \frac{1}{\rho_0} \mathrm{Div}(\bs{q}_0) = r_{\Theta} + \bs{M}^{\mathrm{th}} \colon \bs{L}^{\mathrm{th}} + \bs{M}^{\mathrm{i}} \colon \bs{L}^{\mathrm{i}} + \bs{H}_{\gamma} \colon \dot{\bs{\gamma}} + \bs{H}_F \colon \dot{\bs{F}}^{\mathrm{e}} \text{,} 
\label{hce1}\\
& c_d := - \Theta \frac{\partial^2 \Psi}{\partial \Theta^2} \text{ ,} \quad \bs{H}_{\gamma} := \bs{q}_{\gamma} + \Theta \frac{\partial^2 \Psi}{\partial \Theta \, \partial \bs{\gamma}} \text{ ,} \quad  \bs{H}_F := \Theta \frac{\partial^2 \Psi}{\partial \Theta \, \partial \bs{F}^{\mathrm{e}}} \text{ .}
\label{hce2}
\end{align}

The standard notation introduced in \eqref{hce2} includes  the heat capacity $c_d$,  the~latent heat due to internal microstructural processes in the material $\bs{H}_{\gamma}$ and the latent heat due to deformations $\bs{H}_F$. The~latter enables, amongst others, the~modeling of the Gough--Joule effect~\cite{MCBRIDE20112116}.  

\section{Thermomechanical Modeling of the~SIC}
\label{thmemo}

Whereas Section \ref{basconc} provides the general framework for modeling thermomechanical processes, Section \ref{thmemo} presents the assumptions specific to the modeling of the SIC. This includes the choice of internal variables, the~assumption for the Helmholtz free energy, the~definition of coupling conditions, and the formulation of the dissipation~potential.

\subsection{Definition of Internal~Variables}
\label{intvar}
The multiplicative decomposition of the deformation gradient  representing the  base of kinematics in the finite thermomechanical framework (Equation~\eqref{defgrad}) already incorporates two internal variables, namely the inelastic deformation gradient $\bs{F}^{\mathrm{i}}$ and its thermal counterpart $\bs{F}^{\mathrm{th}}$. In~the case of the SIC, the~inelastic deformations are caused by the crystallization, which will be  emphasized by introducing index ``c'' instead of index ``i'' in the following text (i.e., $\bs{F}^{\mathrm{c}} := \bs{F}^{\mathrm{i}}$,  $\bs{L}^{\mathrm{c}} := \bs{L}^{\mathrm{i}}$, $\bs{M}^{\mathrm{c}} := \bs{M}^{\mathrm{i}}$ and $\mathcal{D}^{\mathrm{c}} := \mathcal{D}^{\mathrm{i}}$ ).

However, the~description of microstructural processes related to SIC requires the introduction of two additional internal variables: regularity of polymer chain network ($\chi$) and thermal flexibility of polymer network ($\alpha$). The~regularity in this context comprises the information on the ordering of the polymer chains with respect to each other and the degree of order among the polymer atoms. It~takes values from the range $[0, 1]$, such that values close to zero correspond to an amorphous state, whereas values close to the value of one are classified as crystalline regions. During~a tensile test, the~regularity evolves, thus simulating the formation/degradation of crystalline regions. The~second internal variable captures physical properties related to the molecular mobility in the polymer which is governed by the chain flexibility and the temperature~\cite{BRADY2017181}. The~flexibility of the polymer chains in the network changes due to warming up and cooling during a cyclic tensile test, where heating induces the chains to become more~flexible.

\subsection{Assumption for the Helmholtz Free Energy~Density}
\label{henergy}
{
In order to obtain a model suitable for simulating the SIC, the~Helmholtz free energy is assumed to consist of four terms combining the benefits of several modeling approaches
\begin{equation}
\Psi = \Psi^{\mathrm{e}}(\bar{I}_1, J^{\mathrm{e}}) +  \Psi^{\mathrm{th}} (\Theta) + \Psi^{\mathrm{crys}} (\chi)  + \Psi^{\mathrm{coup}}(\alpha, \Theta) \text{ .}
\label{psitot}
\end{equation}
 
Here, the~elastically stored energy $\Psi^{\mathrm{e}}$ (Equation \eqref{psivoldev}) includes the Arruda--Boyce model (Equation~\eqref{ab})  and the volumetric contribution (Equation \eqref{simoa}). 
A~similar combination has been used in works by Gasser and Holzapfel (2002) \cite{gasser2002} and by Elguedj and Hughes (2014) \cite{ELGUEDJ2014388} when studying nearly incompressible materials behavior. The~thermally stored energy part $\Psi^{\mathrm{th}}$ has been proposed by Raniecki and Bruhns (1991)~\cite{Raniecki1991} and has recently been used by Mahnken (2013)~\cite{MAHNKEN20132003} for thermomechanical modeling of polymers
\begin{equation}
 \Psi^{\mathrm{th}} (\Theta) = c_d \, \left( \Theta - \Theta_0 \right) - c_d \, \Theta \, \ln\left( \frac{\Theta}{\Theta_0} \right) \text{ .} 
\end{equation}

The thermal contribution above is a function of reference temperature $\Theta_0$ and holds under the condition that the heat capacity $c_d$ is independent of temperature.
The crystalline energy part $\Psi^{\mathrm{crys}}$ is proposed in~\cite{AYGUN2020129}  to model the regularity evolution during the unloading phase. It is a linear function
\begin{equation}
\Psi^{\mathrm{crys}} (\chi) = c_1 \, \chi \text{ ,}
  \label{psicrys}
\end{equation}
where parameter $c_1$ is load-dependent and enables to control the evolution direction. The~last energy contribution,  $\Psi^{\mathrm{coup}}$, is a mixed term in thermal flexibility and temperature, including proportionality constant $c_2$
\begin{equation}
\Psi^{\mathrm{coup}}(\alpha, \Theta) = c_2 \, \alpha \left( \Theta - \Theta_0 \right) \text{ .}
\label{psicoup}
\end{equation}

A similar bilinear form  is suggested in the work by Hajidehi and Stupkiewicz (2017)~\cite{Hajidehi2017}  for the energy transition in shape memory alloys with the aim to simulate the temperature  decrease during~unloading.
}%

\subsection{Coupling~Conditions}
\label{couplcond}
The  problem formulation presented  depends on four internal variables, two of which have a tensorial character. However, some additional information is provided by the physics of the SIC phenomenon which enables the simplification of the underlying setup. The~focus is first set on the velocity gradient $\bs{L}^{\mathrm{c}} $ referring to the rate of deformations caused by the crystallization. The~closer specification of this quantity relies on the experimental observation, showing that the orientation of crystalline regions is dependent on the external load orientation.  This  implies that velocity gradient $\bs{L}^{\mathrm{c}} $ is coaxial with the Mandel stress tensor $\bs{M}^{\mathrm{c}}$ since they are a thermodynamically conjugated pair. Furthermore, it can be expected that the intensity of deformations due to the crystallization depends on the regularity degree of network, which leads to the following assumption for the velocity gradient
\begin{equation}
\bs{L}^{\mathrm{c}} = k_1 \, \dot{\chi} \, \frac{\bs{M}^{\mathrm{c, dev}}}{\lVert \bs{M}^{\mathrm{c, dev}} \rVert} \text{ ,} 
\qquad 
\bs{M}^{\mathrm{c, dev}}  = \bs{M}^{\mathrm{c}} - \frac{\mathrm{tr}(\bs{M}^{\mathrm{c}})}{3} \bs{I} \text{ .}
  \label{coup1}
\end{equation}

Here, $k_1$ denotes the proportionality constant and the choice of the  deviatoric part of the Mandel stress tensor is substantiated by the fact that the model deals with the unfilled polymers being the representatives   of nearly incompressible materials.
In a similar way, the~thermal velocity gradient and the rate of thermal flexibility are coupled to each other through the linear relationship with the  positive proportionality constant $k_2$
\begin{equation}
  \bs{L}^{\mathrm{th}} = k_2 \, \dot{\alpha} \, \bs{I} \text{ .}
  \label{coup2}
\end{equation}

Equation~\eqref{coup2} suggests a diagonal form of $\bs{L}^{\mathrm{th}}$ since an isotropic character of thermal  effects is expected. 
After introducing coupling conditions \eqref{coup1} and \eqref{coup2}, the~problem formulation only depends on the two scalar internal variables $\chi$ and $\alpha$.

\subsection{Derivation of Driving~Forces}
\label{drfor}
Assumptions made in Sections \ref{henergy} and \ref{couplcond} enable the specification of driving forces corresponding to remaining internal variables $\chi$ and $\alpha$.
For this purpose, coupling conditions \eqref{coup1} and \eqref{coup2} are first introduced into the dissipation inequality \eqref{di3a}
\begin{align}
  \mathcal{D} &= \bs{M}^{\mathrm{th}} \colon \bs{L}^{\mathrm{th}} + \bs{M}^{\mathrm{c}} \colon \bs{L}^{\mathrm{c}} + q_{\chi} \, \dot{\chi} + q_{\alpha} \, \dot{\alpha} + \mathcal{D}^{\mathrm{cond}}
  \label{di4} \\
  &= k_2 \, \mathrm{tr}(\bs{M}^{\mathrm{th}}) \, \dot{\alpha} + k_1 \, \left\lVert \bs{M}^{\mathrm{c, dev}} \right\rVert \dot{\chi} + q_{\chi} \, \dot{\chi} + q_{\alpha} \, \dot{\alpha} + \mathcal{D}^{\mathrm{cond}} \text{ .}
  \label{di5}
\end{align}

Thereafter,  $q_{\chi} $   and $q_{\alpha}$ are calculated according to the standard definitions applied to the  assumptions on energy contributions  \eqref{psicrys} and \eqref{psicoup}
\begin{equation}
  q_{\chi} = - \frac{\partial \Psi^{\mathrm{crys}}}{\partial \chi} = - c_1 \text{ ,} 
	\qquad 
	q_{\alpha} = - \frac{\partial \Psi^{\mathrm{coup}}}{\partial \alpha} = - c_2 \left( \Theta - \Theta_0 \right) \text{ .}
	\label{drivf1}
\end{equation}

Now, the~implementation of Equation~\eqref{drivf1} in \eqref{di5} and gathering of terms including the same type of rates yield expressions for sought driving forces
\begin{align}
  &\widetilde{q}_{\chi} := k_1 \, \left\lVert \bs{M}^{\mathrm{c, dev}} \right\rVert - c_1 \text{ ,} 
  \label{drivfa}\\
  &\widetilde{q}_{\alpha} := k_2 \, \mathrm{tr}( \bs{M}^{\mathrm{th}} ) - c_2 \left( \Theta - \Theta_0 \right) \text{ ,}
  \label{drivfb}
\end{align}
such that the dissipation inequality turns into
\begin{equation}
  \mathcal{D} = \widetilde{q}_{\chi} \, \dot{\chi} + \widetilde{q}_{\alpha} \, \dot{\alpha} + \mathcal{D}^{\mathrm{cond}} \geq 0 \text{ .}
  \label{di6}
\end{equation}

The requirement for the non-negativity of each addend yields the conclusion that a driving force must have the same sign as the rate of the corresponding internal variable. This furthermore implies that a decrease of the regularity ($\dot{\chi}<0$) is accompanied by a negative driving force ($\widetilde{q}_{\chi}<0$) during the unloading stage. It is furthermore assumed that the rate of driving force~\eqref{drivfa}
\begin{equation}
  \dot{\widetilde{q}}_{\chi} = k_1 \, \dot{\overline{\left\lVert \bs{M}^{\mathrm{c, dev}} \right\rVert}}
  \label{dotdrivf}
\end{equation}
distinguishes the loading stage ($\dot{\widetilde{q}}_{\chi} \geq 0$) and the unloading stage ($\dot{\widetilde{q}}_{\chi} < 0$). 

\subsection{Assumption for the Dissipation Potential and Derivation of Evolution~Equations}
\label{assumdiss}
The assumed dissipation potential includes two contributions, each of them dependent on the rate of a single internal variable
\begin{equation}
  \Delta(\dot{\chi},\dot{\alpha}) = \Delta^{\mathrm{crys}} (\dot{\chi}) + \Delta^{\mathrm{th}}(\dot{\alpha}) \text{ .}
  \label{disspot}
\end{equation}

The choice of the potential part  related to the network regularity  $\Delta^{\mathrm{crys}}$ is motivated by the experimental results shown in  Figure~\ref{fig1}a. According to this diagram, the~crystalline regions start to evolve after exceeding a threshold value and grow up to the end of the loading phase where the limiting stretch of $\approx 600 \%$ must not be exceeded. Subsequently, the~crystalline regions shrink during the unloading with a lower rate. The~dissipation potential  suitable for modeling this process has been proposed in~\cite{AYGUN2020129}  and has the following form
\begin{equation}
\label{deltacrys}
  \Delta^{\mathrm{crys}} (\dot{\chi}) = \left( A + B \right) \left\vert \dot{\chi} \right\vert \text{ .}
\end{equation}

Here, $A$ is a crystallization limit for the evolution of the regularity and $B$ represents the increment of this limit. The~rate of increment $B$  is a function of the regularity rate
\begin{equation} 
  \dot{B} = \frac{b}{f(\chi)} \left\vert \dot{\chi} \right\vert \text{ ,} \quad f(\chi) = \beta_1 - \left( \chi - \beta_2 \right)^{\beta_3} \text{ ,} \quad b = \begin{cases}
    b_1 & \text{if} \quad \dot{\tilde{q}}_{\chi} \geq 0 \text{ ,} \\
    b_2 & \text{if} \quad \dot{\tilde{q}}_{\chi} < 0 \text{ .}
  \end{cases} 
  \label{dotB}
\end{equation}

The chosen formulation depends on  material constants $\beta_1$, $\beta_2$, and $\beta_3$ and introduces function $f>0$  to favor the evolution of the regularity in regions with higher values.  Moreover, the~load-dependent parameter $b$ controls the velocity of the evolution. The~values for this parameter are different for the loading phase ($\dot{\tilde{q}}_{\chi} \geq 0$) and unloading phase ($\dot{\tilde{q}}_{\chi} < 0$). Condition  $b_2 > b_1 > 0$  indicates a slower decrease of the regularity during the unloading phase than its growth during the loading~phase.

 The second part of the dissipation potential, $\Delta^{\mathrm{th}}$,  is defined in terms of the rate of thermal flexibility and depends on material constants $D_1$ and $D_2$
\begin{equation}
  \Delta^{\mathrm{th}} (\dot{\alpha}) = \frac{D_1}{2} \chi^{-D_2} \, \dot{\alpha}^2 \text{ .}
  \label{pot_alpha}
\end{equation}

The proposed dissipation potentials now serve as a basis for the application of the minimum principle (Section~\ref{mindiss}). To~this end, the~notation introduced in Section~\ref{couplcond} is used to reformulate the rate of Helmholtz energy $\dot{\Psi}$ (Equation~\eqref{dotpsi}) and the Lagrange function $\mathcal{L}$ (Equation~\eqref{min2})  

\begin{gather}
  \dot{\Psi} = \bs{P} \colon \dot{\bs{F}} - \eta \, \dot{\Theta} - \widetilde{q}_{\chi} \, \dot{\chi} - \widetilde{q}_{\alpha} \, \dot{\alpha} \text{ ,}
\\[1mm]
  \mathcal{L} = \bs{P} \colon \dot{\bs{F}} - \eta \, \dot{\Theta} - \widetilde{q}_{\chi} \, \dot{\chi} - \widetilde{q}_{\alpha} \, \dot{\alpha} +  \left( A + B \right) \left\vert \dot{\chi} \right\vert + \frac{D_1}{2} \chi^{-D_2} \, \dot{\alpha}^2 \text{ .}
  \label{lagr}
\end{gather}

Subsequently, the~minimization procedure yields the modified driving forces
\begin{align}
  &\widetilde{q}_{\chi} = \frac{\partial \Delta^{\mathrm{crys}}}{\partial \dot{\chi}} \text{ ,} 
  \label{moddrfora}\\
  &\widetilde{q}_{\alpha} = \frac{\partial \Delta^{\mathrm{th}}}{\partial \dot{\alpha}} \text{ .}
  \label{moddrforb}
\end{align}

However, the~evaluation of the driving force $ \widetilde{q}_{\chi} $ (Equation~\eqref{moddrfora}) is not straightforward since potential $\Delta^{\mathrm{crys}}$ (Equation~\eqref{deltacrys}) depends on the absolute value function which is not differentiable at $\dot{\chi} = 0$. For~that reason, the~subdifferential of the crystalline dissipation potential $\partial \Delta^{\mathrm{crys}}(\dot{\chi})$ is introduced instead of its~derivative

\begin{empheq}[left={\partial \Delta^{\mathrm{crys}} (\dot{\chi}) = \empheqlbrace}]{align}
    \widetilde{q}_{\chi} = \left( A + B \right) \frac{\dot{\chi}}{\left\vert\dot{\chi} \right\vert}  & \quad \text{for} \quad \dot{\chi} \neq 0 \text{ ,}
    \label{sd_a}\\
    \left\{ \widetilde{q}_{\chi}; \, \left\vert \widetilde{q}_{\chi} \right\vert \le A + B \right\} & \quad \text{for} \quad \dot{\chi} = 0 \text{ .}
    \label{sd_b}
\end{empheq}

Equation~\eqref{sd_a} represents the derivative of the potential with respect to the regularity rate under the condition that $\dot{\chi} \neq 0$, whereas any value $\left\vert \widetilde{q}_{\chi} \right\vert \le A + B$ can be a solution for $\dot{\chi} = 0$. Case~\eqref{sd_b} applies at the beginning of the tensile test where the material undergoes purely elastic deformations without changing the regularity. After~exceeding the crystallization limit, the~regularity starts to increase according to the evolution law which is obtained  by rearranging Equation~\eqref{sd_a}
\begin{equation}
  \dot{\chi} = \frac{\left\vert\dot{\chi} \right\vert}{A + B} \widetilde{q}_{\chi} = \lambda \, \widetilde{q}_{\chi} \text{ ,} \quad \lambda \geq 0 \text{ .}
  \label{dotchi}
\end{equation}

The procedure for determining parameter $\lambda$ has been comprehensively explained in~\cite{AYGUN2020129}.
 Within~the present contribution, only the main steps are summarized as follows:
 First, the~absolute values are taken from both sides of the Equation~\eqref{sd_a} and the obtained relationship is squared. In~a second step, after~taking the time derivative of modified Equation~\eqref{sd_a} and inserting Equations~\eqref{drivfa}, \eqref{dotdrivf}, \eqref{dotB} and~\eqref{dotchi}, the~crystallization parameter turns into
\begin{equation}
  \lambda = \frac{f \, \widetilde{q}_{\chi} \, \dot{\widetilde{q}}_{\chi}}{b \left(A + B \right)^2} \ge 0 \text{ .}
  \label{crystpar}
\end{equation}

Equation~\eqref{sd_a} also holds for the complete unloading stage, where the degradation of crystalline regions occurs. This is achieved by introducing a shift  yielding a negative driving force $\widetilde{q}_{\chi}$ during the unloading (Section~\ref{discunload}). Simultaneously, the~rate $\dot{\widetilde{q}}_{\chi}$ is negative per definition during the unloading stage (Equation~\eqref{dotdrivf}). Since driving force $\widetilde{q}_{\chi}$ and its rate $\dot{\widetilde{q}}_{\chi}$ are both negative during the unloading phase, the~condition $\lambda \ge 0$  (Equation\eqref{crystpar}) holds and relationship \eqref{dotchi} applies.

The study of thermal influences accompanying crystallization are based on Equation~\eqref{moddrforb}
 yielding the following evolution equation
\begin{equation}
  \widetilde{q}_{\alpha} = \frac{\partial \Delta^{\mathrm{th}}}{\partial \dot{\alpha}} = D_1 \, \chi^{- D_2} \, \dot{\alpha} 
\quad \Rightarrow \quad
  \dot{\alpha} = \frac{\chi^{D_2}}{D_1} \widetilde{q}_{\alpha} \text{ .}
  \label{dotalpha}
\end{equation}

Obviously, the~rate $\dot{\alpha}$ is proportional to its driving force which characterizes the viscous type of evolution. In~this case, the~decrease of the temperature during the unloading phase is controlled by a suitable choice of constant $c_2$ in the driving force (Equation~\eqref{drivfb}).

Finally, the~insertion of evolution equations~\eqref{dotchi} and \eqref{dotalpha} in \eqref{di6} yields the single dissipation~contributions
\begin{equation}
  \mathcal{D}^{\chi} = \widetilde{q}_{\chi} \, \dot{\chi} = \lambda \, \widetilde{q}_{\chi}^2 \ge 0 \text{ ,} 
	\qquad 
	\mathcal{D}^{\alpha} = \widetilde{q}_{\alpha} \, \dot{\alpha} = \frac{\chi^{D_2}}{D_1} \widetilde{q}_{\alpha}^2 \ge 0 \text{ ,}
  \label{di7}
\end{equation}
where both terms fulfill the non-negativity requirement, which approves the thermodynamical consistency of the~model.

\section{Numerical Implementation of the SIC~Model}
\label{numimpl}
\vspace{-6pt}

\subsection{Implementation of the Thermomechanical Coupled Problem into the~FEM}
\label{implfem}
Balance equations~\eqref{balance1} and \eqref{hce1}  are numerically solved by applying the FEM where a setup appropriate for nonlinear materials and large deformations is developed. This includes the transformation of the energy balance and the development of the weak form of the problem.

The balance of energy \eqref{hce1} is more closely specified by using driving forces (Equations~\eqref{drivfa} and \eqref{drivfb}) along with coupling conditions \eqref{coup1} and \eqref{coup2}
\begin{equation}
\label{benergy}
  c_d \, \dot{\Theta} + \frac{1}{\rho_0} \mathrm{Div}(\bs{q}_0) = r_{\Theta} + \widetilde{H}_{\chi} \, \dot{\chi} + \widetilde{H}_{\alpha} \, \dot{\alpha} + \bs{H}_F \colon \dot{\bs{F}}^{\mathrm{e}} \text{ .}
\end{equation}

Here, latent heat $\widetilde{H}_{\chi}$ and $\widetilde{H}_{\alpha}$ are expressed according to the definition \eqref{hce2}
\begin{equation}\label{lat_heat_1}
  \widetilde{H}_{\chi} := \widetilde{q}_{\chi} + \Theta \frac{\partial^2 \Psi}{\partial \Theta \, \partial \chi} \text{ ,} 
	\qquad 
	\widetilde{H}_{\alpha} := \widetilde{q}_{\alpha} + \Theta \frac{\partial^2 \Psi}{\partial \Theta \, \partial \alpha} \text{ }
\end{equation}
whereas $\bs{H}_F $ remains  as defined in Equation~\eqref{hce2}.
Furthermore, the~energy derivatives in the latent heat contributions are calculated on the basis of assumptions~\eqref{psitot}--\eqref{psicoup}
\begin{equation}\label{lat_heat_2}
  \widetilde{H}_{\chi} = \widetilde{q}_{\chi} \text{ ,} \quad \widetilde{H}_{\alpha} =  \widetilde{q}_{\alpha} + \Theta \, c_2 \text{ ,} \quad \bs{H}_F = \bs{0} \text{ .}
\end{equation}

Provided that the maximal cooling of natural rubber due to the Gough--Joule effect is about \mbox{$0.005$ K~\cite{holzapfel1996}}, the~temperature change by latent heat $\bs{H}_F$ is negligible compared to the temperature change due to the SIC. This motivates the choice of energy density $\Psi$ (Equation \eqref{psitot}), where it holds $\frac{\partial^2\Psi}{\partial\Theta \partial\bs{F}^{\mathrm{e}}}=\bs{0}$, and thus the corresponding latent heat contribution becomes $\bs{H}_F=\bs{0}$. Accordingly, balance~of energy \eqref{benergy} turns into
\begin{equation}
  c_d \, \dot{\Theta} + \frac{1}{\rho_0} \mathrm{Div}(\bs{q}_0) = r_{\Theta} + \widetilde{q}_{\chi} \, \dot{\chi} + \left( \widetilde{q}_{\alpha} + \Theta \, c_2 \right) \dot{\alpha} \text{ .}
  \label{balance3}
\end{equation}

For the purpose of transforming the strong formulation of the problem into its  weak form, two~steps are performed: First, the~strong form of each differential equation (Equations~\eqref{balance1} and \eqref{balance3}) is multiplied by test functions $\delta \bs{u}$ and $\delta \Theta$, commonly referred to as virtual displacement and virtual temperature. Thereafter, the~equations are integrated over domain $\mathcal{B}$. Finally, the~integration by parts and subsequently the Gauss integration theorem are applied to the divergence term in Equation~\eqref{balance3} to transform the volume integral over the body $\mathcal{B}$ into an integral over the surface $\partial \mathcal{B}$. The~weak form of both balance equations reads
\begin{equation}
  \int_{\mathcal{B}} \nabla_{\bs{X}} \delta \bs{u} \colon \bs{P} \, \mathrm{d} V - \int_{\mathcal{B}} \delta \bs{u} \cdot \rho_0 \, \bs{b} \, \mathrm{d}V - \int_{\partial \mathcal{B}^{\bs{t}}} \delta \bs{u} \cdot \bs{t} \, \mathrm{d} A = 0 \text{ ,}
\end{equation}
\begin{equation}
\begin{split}
& \int_{\mathcal{B}} \delta \Theta \, c_d \, \dot{\Theta} \, \mathrm{d} V + \int_{\partial \mathcal{B}} \delta \Theta \frac{1}{\rho_0} \bs{q}_0 \cdot \bs{n} \, \mathrm{d} A - \int_{\mathcal{B}} \nabla_{\bs{X}} \delta \Theta \cdot \frac{1}{\rho_0} \bs{q}_0 \, \mathrm{d} V - \\
& \int_{\mathcal{B}} \delta \Theta \, r_{\Theta} \, \mathrm{d} V - \int_{\mathcal{B}} \delta \Theta \, \widetilde{q}_{\chi} \, \dot{\chi} \, \mathrm{d} V - \int_{\mathcal{B}} \delta \Theta \left( \widetilde{q}_{\alpha} + \Theta \, c_2 \right) \dot{\alpha} \, \mathrm{d} V = 0 \text{ .}
\end{split}
\end{equation}

In a second step, domain $\mathcal{B} \approx \bigcup_{e=1}^{n_{\mathrm{el}}} \mathcal{B}^{e}$ is spatially disrcretized into a finite number of elements $n_{\mathrm{el}}$ and integrals are approximated by a sum of integrals over single elements $\mathcal{B}^{e}$. The~same applies for the boundary parts:
\begin{equation}
 \sum_{e = 1}^{n_{\mathrm{el}}} \left[ \int_{\mathcal{B}^{e}} \nabla_{\bs{X}} \delta \bs{u}^{e} \colon \bs{P} \, \mathrm{d} V - \int_{\mathcal{B}^{e}} \delta \bs{u}^{e} \cdot \rho_0 \, \bs{b} \, \mathrm{d}V - \int_{\partial \mathcal{B}^{e \, \bs{t}}} \delta \bs{u}^{e} \cdot \bs{t} \, \mathrm{d} A \right] = 0 \text{ ,}
 \label{dwf_1}
\end{equation}
\begin{equation}
\begin{split}
& \sum_{e=1}^{n_{\mathrm{el}}} \biggl[ \int_{\mathcal{B}^{e}} \delta \Theta^{e} \, c_d \, \dot{\Theta}^{e} \, \mathrm{d} V + \int_{\partial \mathcal{B}^{e}} \delta \Theta^{e} \frac{1}{\rho_0} \bs{q}_0 \cdot \bs{n} \, \mathrm{d} A - \int_{\mathcal{B}^{e}} \nabla_{\bs{X}} \delta \Theta^{e} \cdot \frac{1}{\rho_0} \bs{q}_0 \, \mathrm{d} V - \\
& \phantom{\sum_{e=1}^{n_{\mathrm{el}}} \biggl[} \int_{\mathcal{B}^{e}} \delta \Theta^{e} \, r_{\Theta} \, \mathrm{d} V - \int_{\mathcal{B}^{e}} \delta \Theta^{e} \, \widetilde{q}_{\chi} \, \dot{\chi} \, \mathrm{d} V - \int_{\mathcal{B}^{e}} \delta \Theta^{e} \left( \widetilde{q}_{\alpha} + \Theta \, c_2 \right) \dot{\alpha} \, \mathrm{d} V \biggr] = 0 \text{ .}
\end{split}
\label{dwf_2}
\end{equation}

The field variables $\delta \bs{u}^{e}$ and $\delta \Theta^{e}$ and their gradients $\nabla_{\bs{X}} \delta \bs{u}^{e}$ and $\nabla_{\bs{X}} \delta \Theta^{e}$ refer to single elements and are  interpolated by using an approximation with $n_{\mathrm{en}}$ support points
\begin{align}
  & \delta \bs{u}^{e} \approx \sum_{A=1}^{n_{\mathrm{en}}} \delta \bs{u}^{e A} \, N^A \text{ ,} \quad \delta \Theta^{e} \approx \sum_{A=1}^{n_{\mathrm{nen}}} \delta \Theta^{e A} \, N^A \text{ ,} 
  \label{interp1} \\
  & \nabla_{\bs{X}} \delta \bs{u}^{e} \approx \sum_{A=1}^{n_{\mathrm{en}}} \delta \bs{u}^{e A} \otimes \nabla_{\bs{X}} N^A \text{ ,} \quad \nabla_{\bs{X}} \delta \Theta^{e} \approx \sum_{A=1}^{n_{\mathrm{nen}}} \delta \Theta^{e A} \, \nabla_{\bs{X}} N^A \text{ ,}
  \label{interp2}
\end{align}
where $\delta \bs{u}^{e A}$ and $\delta \Theta^{e}$ are the values at node $A$ of element $e$. The~interpolation is done using C$^0$-continuous Lagrange basis functions $N^A$. The~application of Equations~\eqref{interp1} and \eqref{interp2} on the discretized weak forms (Equations~\eqref{dwf_1} and \eqref{dwf_2}) yields
\begin{align}
  & \sum_{e = 1}^{n_{\mathrm{el}}} \left\{  \sum_{A=1}^{n_{\mathrm{en}}} \delta \bs{u}^{e A} \cdot \left[ \bs{f}_{u, \mathrm{int}}^{e A} - \bs{f}_{u, \mathrm{vol}}^{e A} - \bs{f}_{u, \mathrm{sur}}^{e A} \right] \right\} = 0 \text{ ,} 
  \label{dwf_contr1}\\
  & \sum_{e = 1}^{n_{\mathrm{el}}} \left\{  \sum_{A=1}^{n_{\mathrm{nen}}} \delta \Theta^{e A} \left[ f_{\Theta, \mathrm{trans}}^{e A} + f_{\Theta, \mathrm{cond}}^{e A} - f_{\Theta, \mathrm{vol}}^{e A} - f_{\Theta, \mathrm{sur}}^{e A} - f_{\Theta, \mathrm{crys}}^{e A} - f_{\Theta, \mathrm{flex}}^{e A} \right] \right\} = 0 \text{ .}
  \label{dwf_contr2}
\end{align}

The force contributions in the discretized weak form of the balance of linear momentum (Equation~\eqref{dwf_contr1}) correspond to the internal forces $\bs{f}_{u, \mathrm{int}}^{e A}$, the~volumetric forces $\bs{f}_{u, \mathrm{vol}}^{e A}$, and the surface forces $\bs{f}_{u, \mathrm{sur}}^{e A}$:
\begin{align}
  & \bs{f}_{u, \mathrm{int}}^{e A} := \int_{\mathcal{B}^{e}} \bs{P} \cdot \nabla_{\bs{X}} N^A \, \mathrm{d} V
\text{ ,} \quad \bs{f}_{u, \mathrm{vol}}^{e A} := \int_{\mathcal{B}^{e}} \rho_0 \, N^A \, \bs{b} \, \mathrm{d}V \text{ ,} \label{f1}\\
  & \bs{f}_{u, \mathrm{sur}}^{e A} := \int_{\partial \mathcal{B}^{e \, \bs{t}}} N^A \, \bs{t} \, \mathrm{d} A \text{ ,} \label{f2}
\end{align}
whereas, the~discretized weak form of the balance of energy (Equation~\eqref{dwf_contr2}) includes fluxes $f_{\Theta, \mathrm{tran}}^{e A}$ describing transient heat transfer, fluxes $f_{\Theta, \mathrm{cond}}^{e A}$ including heat conduction, volumetric fluxes $f_{\Theta, \mathrm{vol}}^{e A}$, surface fluxes $f_{\Theta, \mathrm{sur}}^{e A}$, fluxes $f_{\Theta, \mathrm{crys}}^{e A}$ yielding a temperature change caused by the crystallization, and fluxes $f_{\Theta, \mathrm{flex}}^{e A}$ changing the temperature due to thermal flexibility
\begin{align}
  & f_{\Theta, \mathrm{tran}}^{e A} := \sum_{B=1}^{n_{\mathrm{nen}}} \dot{\Theta}^{e B} \int_{\mathcal{B}^{e}} N^A \, c_d \, N^B \, \mathrm{d} V \text{ ,} \quad f_{\Theta, \mathrm{vol}}^{e A} := \int_{\mathcal{B}^{e}} N^A \, r_{\Theta} \, \mathrm{d} V \text{ ,}
\label{force1}\\
  & f_{\Theta, \mathrm{cond}}^{e A} := - \int_{\mathcal{B}^{e}} \nabla_{\bs{X}} N^A \cdot \frac{1}{\rho_0} \bs{q}_0 \, \mathrm{d} V \text{ ,} \quad f_{\Theta, \mathrm{sur}}^{e A} := - \int_{\partial \mathcal{B}^{e}} N^A \frac{1}{\rho_0} \bs{q}_0 \cdot \bs{n} \, \mathrm{d} A \text{ ,}
\label{force2}\\
  & f_{\Theta, \mathrm{crys}}^{e A} := \int_{\mathcal{B}^{e}} N^A \, \widetilde{q}_{\chi} \, \dot{\chi} \, \mathrm{d} V \text{ ,} \quad f_{\Theta, \mathrm{flex}}^{e A} := \int_{\mathcal{B}^{e}} N^A \left( \widetilde{q}_{\alpha} + \Theta \, c_2 \right) \dot{\alpha} \, \mathrm{d} V \text{ .}
\label{force3}
\end{align}

A special focus lies on the last flux contribution (Equation~\eqref{force3}), which after inserting Equation~\eqref{drivfb} yields
\begin{equation}
\label{tforce}
  f_{\Theta, \mathrm{flex}}^{e A} := \int_{\mathcal{B}^{e}} N^A \left( q_{\alpha} \, \dot{\alpha} + c_2 \, \Theta_0 \, \dot{\alpha} \right) \mathrm{d} V \text{ .}
\end{equation}

The first term in the parenthesis in Equation~\eqref{tforce} is non-negative due to dissipation inequality~\eqref{di4}. However, the~second term represents a product of positive constants $c_2 \, \Theta_0$ with rate $\dot{\alpha}$ which can be negative during the unloading phase. This allows the reduction of temperature in the unloading phase as it is observed in experimental results  (Figure \ref{fig1}b). Finally, the~element contributions are assembled to a global system of equations under the consideration of kinematic compatibility
\begin{align}
  &\begin{bmatrix}
    \delta \bs{u} \\
    \delta \bs{\Theta}
  \end{bmatrix}^T  
  \cdot \, \bs{R} \left( \bs{u}, \bs{\Theta} \right) = 0 \text{ ,} \label{system}\\
  &\bs{R} \left( \bs{u}, \bs{\Theta} \right) = 
  \begin{bmatrix}
    \bs{f}_{u, \mathrm{int}} - \bs{f}_{u, \mathrm{vol}} - \bs{f}_{u, \mathrm{sur}} \\
    \bs{f}_{\Theta, \mathrm{tran}} + \bs{f}_{\Theta, \mathrm{cond}} - \bs{f}_{\Theta, \mathrm{vol}}- \bs{f}_{\Theta, \mathrm{sur}} - \bs{f}_{\Theta, \mathrm{crys}} - \bs{f}_{\Theta, \mathrm{flex}}
  \end{bmatrix} \text{ .}
\label{resi}  
\end{align}

Provided that $\bs{R} = \bs{0}$ holds for all admissible $[\delta \bs{u} \quad \delta \bs{\Theta}]$, problem \eqref{system} can be solved by any solver for a nonlinear system of equations. { Most commonly, the~Newton-Raphson method is applied  for this purpose. This procedure relies on the following iteration rule
\begin{equation} \label{new_rap}
  \Delta \bs{g}_i = - \bs{K}^{-1} \cdot \bs{R} \text{ ,} \quad\qquad \bs{g}^n_{i+1} = \bs{g}^n_i + \Delta \bs{g}_i \text{ .}
\end{equation}

Here, $i$ denotes the Newton iteration counter, $n$ is the current  time step and $\Delta \bs{g}_i = [\Delta \bs{u}_i \quad \Delta \bs{\Theta}_i]$ are the increments of the global variables $ \bs{g}^n_i = [ \bs{u}^n_i \quad \bs{\Theta}^n_i]$ related to the time step $n$. Finally,  $\,\bs{K}=\frac{\partial\bs{R}}{\partial\bs{g}}$ represents the global  stiffness matrix  of the system. 
}

\subsection{Time Discretization and Simulation of the Unloading~Phase}
\label{discunload}
\textls[-15]{Experimental results show that the loading stage is related to the regularity and temperature increase, whereas the degradation of crystalline regions accompanied by the decrease of the temperature occurs during the unloading stage. In~the present model,  evolution equations~\eqref{dotchi} and \eqref{dotalpha} control the process described. Their numerical implementation relies on the time discretization of the change of crystallization limit (Equation~\eqref{dotB}), of~the crystallization parameter (Equation~\eqref{crystpar}) and of both driving forces (Equations~\eqref{drivfa} and \eqref{drivfb}). The~explicit scheme approximating time derivatives by a forward difference quotient is chosen for this purpose:}
\begin{align}
  & \chi_{n+1} = \chi_n + \Delta \lambda \, \widetilde{q}_{\chi \, n} \text{ ,} 
  \label{dis_chi}\\[1mm]
  & \widetilde{q}_{\chi\, n} = k_1 \lVert \bs{M}^{\mathrm{c, dev}}_n \rVert - c_1 \text{ ,}
  \label{dis_qchi} \\[1mm]
  & \Delta \lambda = \frac{\Delta t \, f(\chi_n) \, \widetilde{q}_{\chi \, n} \, k_1 \left( \lVert \bs{M}^{\mathrm{c, dev}}_n \rVert - \lVert \bs{M}^{\mathrm{c, dev}}_{n-1} \rVert \right)}{b \left( A + B_n \right)^2} \text{ ,} \\[1mm]
  & B_{n+1} = B_n + \frac{b}{f(\chi_n)} \left\vert \chi_{n+1} - \chi_n \right\vert \text{ ,} 
  \label{dis_B}\\[1mm]
  & \alpha_{n+1} = \alpha_n + \Delta t \frac{\chi_n^{D_2}}{D_1} \widetilde{q}_{\alpha \, n} \text{ ,} 
  \label{dis_alpha}\\[1mm]
  & \widetilde{q}_{\alpha \, n} = k_2 \, \mathrm{tr}( \bs{M}^{\mathrm{th}}_n ) - c_2 \left( \Theta_n - \Theta_0 \right) \text{ .}\label{dis_qalpha}
\end{align}

Here, subscript $n+1$ denotes values at the current time step and subscript $n$ refers to the values in the previous time step. It is important to point out that the evaluation of the crystalline and thermal Mandel stresses in the above expressions involve the time integration of tensor valued quantities. For~this reason, evolution laws~\eqref{coup1} and \eqref{coup2} are numerically solved by applying the exponential map~\cite{de2011computational} along with definitions~\eqref{Dtha} and \eqref{Dia} 
\begin{align}
  &\bs{F}^{\mathrm{c}}_{n+1} = \mathrm{exp} \left( k_1 \left( \chi_{n+1} - \chi_n \right) \frac{\bs{M}_n^{\mathrm{c, dev}}}{\lVert \bs{M}_n^{\mathrm{c, dev}} \rVert} \right) \cdot \bs{F}^{\mathrm{c}}_n \text{ ,} 
  \label{dis_Fc}\\
  &\bs{F}^{\mathrm{th}}_{n+1} = \mathrm{exp} \left( k_2 \left( \alpha_{n+1} - \alpha_n \right) \bs{I} \right) \cdot \bs{F}^{\mathrm{th}}_n \text{ .}
   \label{dis_Fth}
\end{align}

The method used in the current approach goes back to the definition of the tensor exponential, where the numerical solution is carried out by calculating a finite truncation of the Taylor series. Equation~\eqref{dis_Fth} is an exact solution due to the diagonality of identity $\bs{I}$. The~contribution by Moler and Van Loan (2003)~\cite{moler2003} discusses various approaches to compute the exponential of a second order tensor and compares their applicability and efficiency.

Within the framework of  the numerical implementation, the~simulation of the degradation of crystalline regions deserves special attention. According to Equation \eqref{dotdrivf}, the~evolution direction of the regularity is controlled by the sign of the driving force due to the non-negativity of $\lambda$. In~other words, a~negative driving force during the unloading stage leads to a reduction of the regularity which is achieved by choosing a suitable shift $c_1$ in Equation~\eqref{dis_qchi}. This load-dependent parameter is calculated from the condition for the initial value of driving force $\widetilde{q}_{\chi}^{\mathrm{un, in}}$ to coincide with the negative crystallization limit if increment $B$ is set to zero
\begin{equation}
  \widetilde{q}_{\chi}^{\mathrm{un, in}} = k_1 \, \lVert \bs{M}^{\mathrm{c, dev}}_{\mathrm{end},\mathrm{ld}} \rVert - c_1 = - A \quad \Rightarrow \quad c_1 = A + k_1 \lVert \bs{M}^{\mathrm{c, dev}}_{\mathrm{end},\mathrm{ld}} \rVert \text{ .}
  \label{q_sh}
\end{equation}

Here, $\bs{M}^{\mathrm{c, dev}}_{\mathrm{end},\mathrm{ld}}$ is the deviatoric crystalline Mandel stress tensor at the end of the loading stage, subscript~``ld'' denotes the loading stage, superscript ``un'' denotes the unloading stage, and ``in'' is an initial~value. 

{
\subsection{Algorithmic Aspects of  the SIC Model~Implementation}
\label{algor}
The SIC model is implemented in the FEAP-software that is appropriate for an enhanced FE-analysis of the complex material behavior. A~new element along with a new material subroutine are developed to this end. Their structure and interconnection with the main program are presented in~Figure~\ref{flowchart}.

The flow chart  of the main program (Figure~\ref{flowchart}a) shows that the element subroutine is used at the beginning of the simulations to allocate material properties to each single element and to generate data on the initial  network regularity. Thereafter, the~element subroutine is called to calculate the element residual and stiffness matrix which  are finally assembled  in the global residual $\bs{R}$ and  global stiffness matrix $\bs{K}$. The~latter are used to solve
the nonlinear system of Equations \eqref{system} and \eqref{resi} according to the scheme \eqref{new_rap}. This iterative procedure is performed until the prescribed accuracy is achieved. To~this end, a~tolerance for the energy norm of displacement increments is set to the value 
$1 \times 10^{-12}$. The~applied monolithic scheme solves the mechanical and thermal balance equations simultaneously with respect to the field variables, displacements and temperature. It provides final values of global external variables $\bs{u}^n$ and $\bs{\Theta}^n$ for each time step $n$. The~time iteration is performed on the basis of the Newmark method with the default parameters $\beta = 0.25$ and $\gamma = 0.5$. The~incorporation of temporal aspects is needed in order to simulate the external load in an incremental form and, correspondingly, to~update the solution of the boundary value problem \eqref{balance1}--\eqref{balance_end}.

The element subroutine  includes two essential tasks, namely the generation of the initial network regularity and the evaluation of the element residual and of the element stiffness matrix (Figure~\ref{flowchart}b). The~generation of the initial regularity is physically motivated by the fact that the natural amorphous entangled microstructure of polymer chains includes places with a different cross-linking degree and polymer chain arrangement. Some of these places are especially suitable for the formation of crystalline regions and behave as  nuclei of the crystallization. In~order to model such an initial configuration, the~network regularity is first set to zero  in all elements. In~a second step, a~uniform distribution is applied to generate initial values of the network regularity in the range 
 $[1 \times 10^{-8}, \, 1 \times 10^{-3}]$
as well as to generate the element numbers associated to the particular initial values. Two intrinsic subroutines of the Fortran program language, random$\_$seed and random$\_$number, are used for this purpose. The~volume fraction  of elements with a higher initial value is varied and fitted with respect to the experimental results shown in Figure~\ref{fig1}a. Depending on the initial configuration, the~material model  determines where the crystalline regions develop and how fast they grow.

The second core assignment of the element subroutine is the definition of the evaluation of the residual and of the element stiffness matrix. The~basis for this  step is already provided by definitions of forces and fluxes (Equations \eqref{f1}--\eqref{force3}). Accordingly, the~residual vector corresponding to the element $e$ has the form
\begin{equation}
  \bs{R}^{e} = 
  \begin{bmatrix}
    \bs{R}^{e}_u \\
    \bs{R}^{e}_{\Theta} 
  \end{bmatrix} = 
  \begin{bmatrix}
    \bs{f}_{u, \mathrm{int}}^{e} - \bs{f}_{u, \mathrm{vol}}^{e} - \bs{f}_{u, \mathrm{sur}}^{e} \\
    \bs{f}_{\Theta, \mathrm{tran}}^{e} + \bs{f}_{\Theta, \mathrm{cond}}^{e} - \bs{f}_{\Theta, \mathrm{vol}}^{e} - \bs{f}_{\Theta, \mathrm{sur}}^{e} - \bs{f}_{\Theta, \mathrm{crys}}^{e} - \bs{f}_{\Theta, \mathrm{flex}}^{e}
  \end{bmatrix} \text{ ,}
  \label{residual}
\end{equation}
and corresponding stiffness matrix $\bs{K}^{{e}}$ results from the derivative of the residual \eqref{residual} with respect to the global field variables $\bs{g} = [ \bs{u} \quad \Theta ]^T$
\begin{equation}
  \bs{K}^{e} = \frac{\partial \bs{R}^{e}}{\partial \bs{g}} = 
  \begin{bmatrix}
    \dfrac{\partial\bs{R}^{e}_u}{\partial \bs{u}} & \dfrac{\partial\bs{R}^{e}_u}{\partial \Theta} \\[3mm]
    \dfrac{\partial \bs{R}^{e}_{\Theta}}{\partial \bs{u}} & \dfrac{\partial \bs{R}^{e}_{\Theta}}{\partial \Theta}
  \end{bmatrix} = 
  \begin{bmatrix}
    \dfrac{\partial \bs{f}_{u, \mathrm{int}}^{e}}{\partial \bs{u}} & \bs{0} \\[3mm]
    - \dfrac{\partial \bs{f}_{\Theta, \mathrm{crys}}^{e}}{\partial \bs{u}} - \dfrac{\partial \bs{f}_{\Theta, \mathrm{flex}}^{e}}{\partial \bs{u}} & \dfrac{\partial \bs{f}_{\Theta, \mathrm{tran}}^{e}}{\partial \Theta} + \dfrac{\partial \bs{f}_{\Theta, \mathrm{cond}}^{e}}{\partial \Theta} - \dfrac{\partial \bs{f}_{\Theta, \mathrm{flex}}^{e}}{\partial \Theta}
  \end{bmatrix}   \text{ .}
  \label{stiffness}
\end{equation}

\begin{figure}[H]
  \centering
  \includegraphics[width=\textwidth]{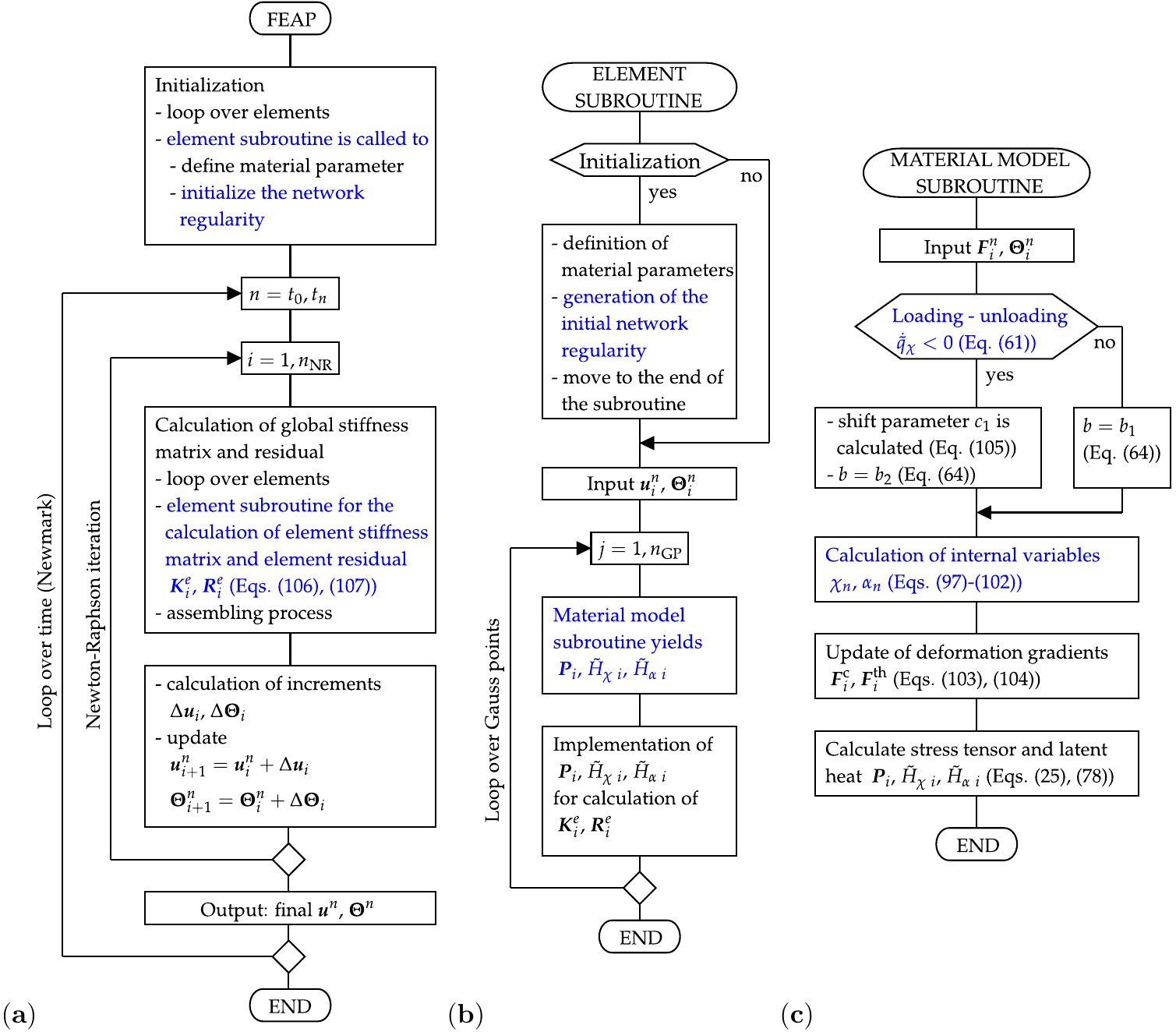}
  \caption{(\textbf{a}) Flowchart of the main program, (\textbf{b}) flowchart of the element subroutine, (\textbf{c}) flowchart of the material model~subroutine.}
\label{flowchart}
\end{figure}

Amongst others, previous equations depend on the stress tensor and on the latent heats which are calculated in the material model subroutine (Figure~\ref{flowchart}c). Here, the~input variables are temperature and deformation gradient. Criterion \eqref{dotdrivf} distinguishes the loading and the unloading mode. Such~a differentiation is needed since the shift $c_1$  (Equation \eqref{q_sh}) and the appropriate value for constant $b$ (Equation~\eqref{dotB}) are defined at the transition from the loading to the unloading mode. Subsequently, three following  tasks are~accomplished: 
\begin{itemize}[leftmargin=13pt,labelsep=7pt]
\item  the update of internal variables according to the Equations \eqref{dis_chi}--\eqref{dis_qalpha}, which also requires the calculation of Mandel stress tensors (Equations \eqref{Dib} and \eqref{Dth2}),
\item the calculations of the crystalline and the thermal deformation gradient are updated by \mbox{Equations \eqref{dis_Fc} and \eqref{dis_Fth}},
\item the calculations of the first Piola--Kirchhoff stress tensor (Equation~\eqref{pi_etaa}), and latent heats (Equation~\eqref{lat_heat_2}). These quantities are final results passed to the element subroutine.
\end {itemize}
}

\section{Numerical~Examples}
\label{results}
\vspace{-6pt}

\subsection{Simulation of Single Crystalline Regions Embedded in the Matrix~Material}
\label{1crys}
The first numerical example deals with the simulation of a tensile test performed on a two-dimensional sample that demonstrates the material behavior under the influence of the SIC. The~elastic material constants are chosen for rubber and correspond to the original amorphous phase~\cite{Shahzad2015, Maeda2015}. The~crystalline material constants are fitted to the experimental results by Toki~et~al. (2003) \cite{toki2003} and Candau~et~al. (2015) \cite{CANDAU2015244}. The~latter work  is also used for fitting thermal material constants (Figure \ref{fig1}b). An~overview of all material parameters is given in Table~\ref{mat_par}. The~thermal conductivity is neglected ($\lambda_{\Theta} = 0$), as~the focus lies on the heat generated or absorbed by SIC and not on the flow of heat within the material. The~conductivity of natural rubber is low compared to the other materials and amounts to 0.15 W/(m K)~\cite{doi:10.1063/1.5088299}. Moreover, no external heat sources ($r_{\Theta} = 0$) are applied in the~simulations.

\begin{table}[H]
  \caption{Elastic, crystalline, and thermal material constants used in~simulations.}
  \centering
  \renewcommand*{\arraystretch}{1.1}
  \begin{tabular}{llcc}
	\toprule
	\textbf{Elastic Constants}&&&\\
    \midrule
    Bulk modulus & $K$ & $5 \times 10^8$ & Pa \\
    Shear modulus & $\mu$ & $4 \times 10^5$ & Pa \\
    Limiting network stretch & $\lambda_m$ & 2 & -\\
    \midrule
	\textbf{Crystalline Constants} &&&\\
	\midrule
	Coupling constant & $k_1$ & 
$7 \times 10^{-2}$ & -\\
	Crystallization limit & $A$ & $1 \times 10^5$ & Pa\\
	Hardening constant & $b_1$ &  $1.7 \times 10^5$ & Pa \\
	Softening constant & $b_2$ &  $2 \times 10^5$ & Pa \\
	Constant in function $f$ & $\beta_1$ &  $0.25$ & - \\
	Constant in function $f$ & $\beta_2$ &  $0.5$ & - \\
	Exponent in function $f$ & $\beta_3$ &  2 & -\\
	\midrule
	\textbf{Thermal Constants} &&&\\
	\midrule
	Reference temperature & $\Theta_0$ & 300 & K \\
    Thermal flexibility modulus & $c_2$ & $2 \times 10^5$ & Pa/K \\
	Coupling constant & $k_2$ & $4 \times 10^{-2}$ & - \\
	Thermal flexibility coefficient & $D_1$ & $3.3 \times 10^6$ & Pa \\
	Exponent & $D_2$ & 6 & - \\
	\bottomrule
  \end{tabular}
  \label{mat_par}	
\end{table}

\begin{figure}[H]
 \centering
 \includegraphics[width=\textwidth]{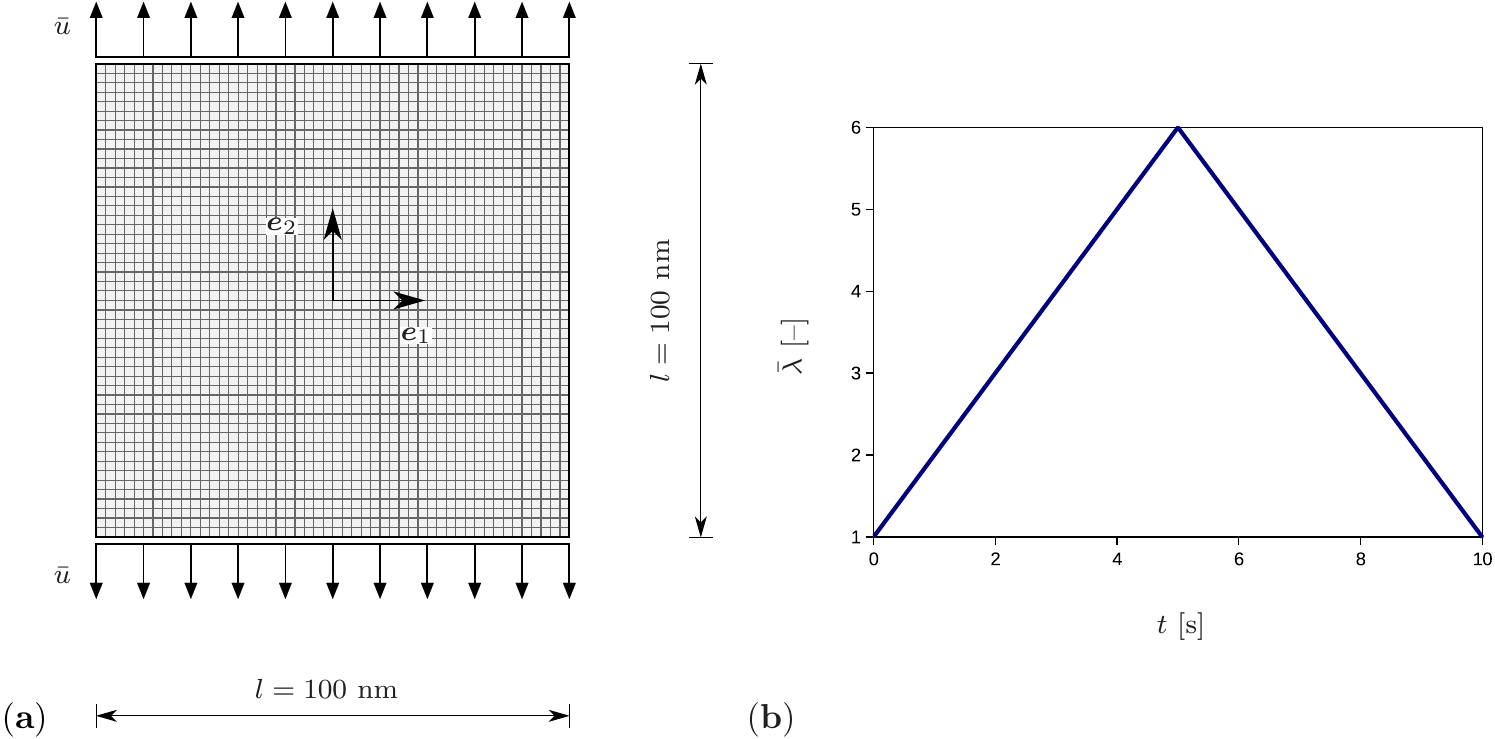} 
 \caption{(\textbf{a}) Geometry and discretization of the sample with the prescribed vertical displacement $\bar{u}$ and side length $l$. (\textbf{b}) The applied stretch $\bar{\lambda} = (l+2\,\bar{u}) / l$ as a function of~time.}
 \label{init}
\end{figure}

The setup corresponding to the tensile test is shown in Figure~\ref{init}a. The~chosen square specimen has the dimensions $100 \times 100$~nm and is discretized by $50 \times 50$ square elements. { For this  purpose, four-node quadrilateral elements with bilinear shape functions are assumed. They are fully integrated by using four integration points. In~addition, the~simulations are performed by applying quadrilateral elements with the reduced one-point integration. However, both types of simulations yield comparable results and no effects of volume locking or of hourglassing are observed.}
The sample thickness is 1~nm and is significantly smaller than the remaining dimensions, which corresponds to a plane stress state problem. The~application of the model to 3D simulations would be straightforward, since the general  SIC material model is presented in previous sections. Vertical displacements prescribed at the horizontal boundaries increase linearly up to the maximum value of 250~nm and thereafter decrease linearly to zero (Figure~\ref{init}b). The~displacement increment is set to $\left\vert \Delta \bar{u} \right\vert = 2.5 \times 10^{-2}$~nm. Here, the~bar symbol indicates external influences. The~prescribed stretch is calculated according to $\bar{\lambda} = (l+2\,\bar{u}) / l$. The~total loading time amounts to 10~s, while the time increment is $\Delta t = 1 \times 10^{-3}$~s.

\begin{figure}[H]
  \centering
  \includegraphics[width=\textwidth]{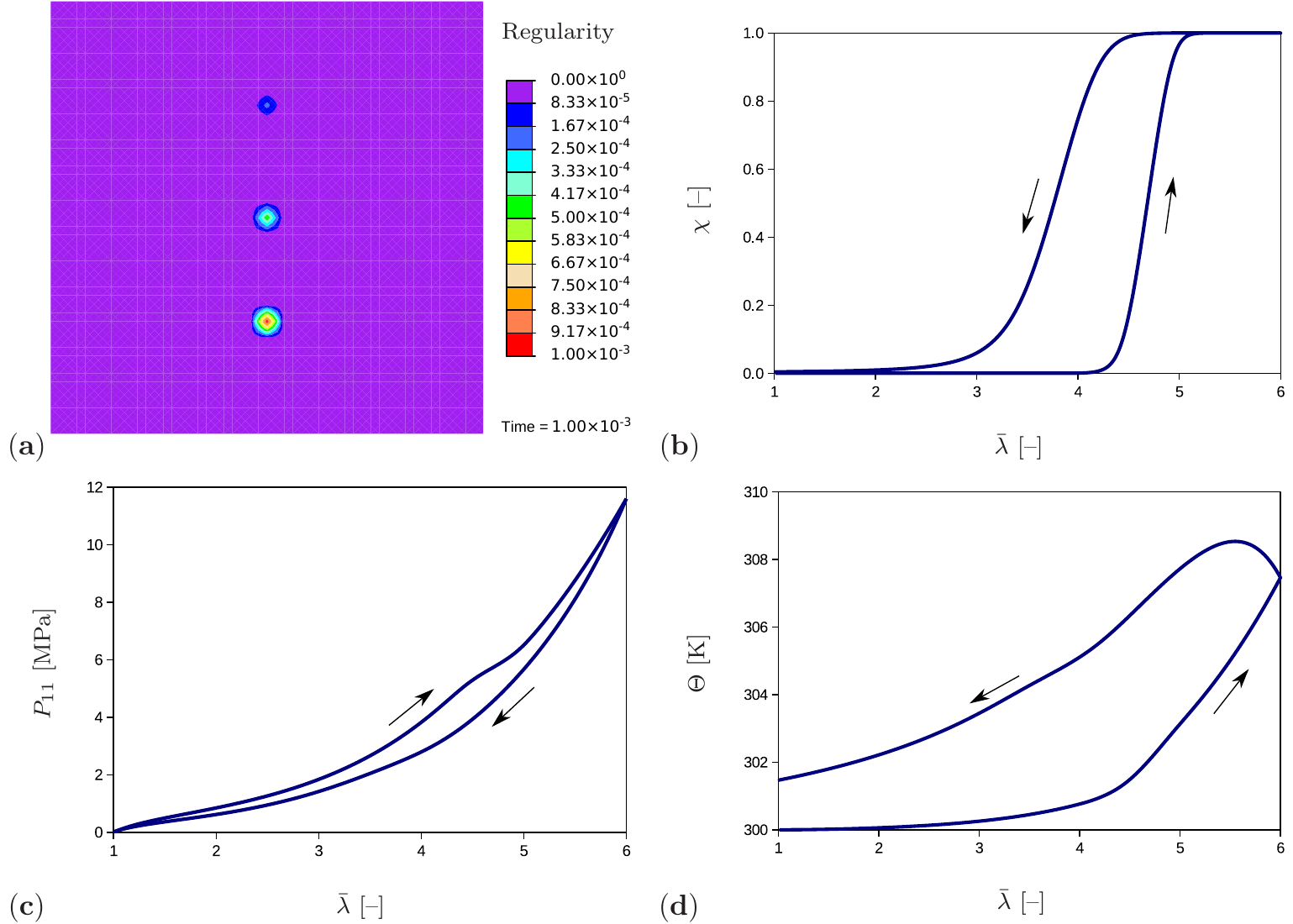}
  \caption{(\textbf{a}) Initial microstructure  
 with three nuclei, (\textbf{b}) the regularity vs. stretch, (\textbf{c}) the stress--stretch diagram, and (\textbf{d}) the temperature change over the stretch at the crystalline region in the bottom part of the~sample.}
  \label{singlecrys1}
\end{figure}

The  example has an academic character and investigates the material response under the presence of three separate crystalline regions embedded in the amorphous matrix material (Figure~\ref{init}). Furthermore, the~influence of the initial value of the regularity degree is studied.
 In the initial configuration (Figure~\ref{singlecrys1}a) the regularity is set to a value of $\chi_0 = 1 \times 10^{-3}$ at an element in the bottom part of the sample, to~$\chi_0 = 1 \times 10^{-4}$ at an element in the middle of the sample and to $\chi_0 = 1 \times 10^{-5}$ at an element in the upper part of the sample. The~evolution of the regularity over the time at the crystalline region in the bottom is shown for the complete load cycle in Figure~\ref{singlecrys1}b. As~expected, the~regularity starts to evolve after exceeding the crystallization limit (Equation~\eqref{sd_b}) and reaches the value of $\chi \approx 1$ corresponding to full crystallization. During~the unloading phase, it decreases with a slower rate (Equation~\eqref{dotB}) and ends up at the value of $\chi \approx 0$ complying with the amorphous material. 
{ 
 A wide range of experimental results show that crystalline regions completely recede at stretches between 250\% and 350\% and that the original amorphous phase is completely recovered. Examples giving evidence of this phenomenon are presented in Le Gac~et~al. (2018) \cite{LEGAC2018209} for  polychloroprene and in Huneau (2011)~\cite{huneau2011} for polyisoprene. One exception is shown in the work by Albouy~et~al. (2005) dealing with the inverse yielding effect. An~incomplete recovery  only occurs if the sample was previously loaded beyond a  certain critical draw ratio. Additional factors that benefit this effect are  a low cross-linking density and  lower temperatures. }
The stress--stretch curve (Figure~\ref{singlecrys1}c) builds a hysteresis loop which is in agreement with the experimental results { (Figure \ref{fig1}a)}. Here, the~$P_{22}$ component of the first Piola--Kirchhoff stress tensor is plotted. The~temperature curve (Figure~\ref{singlecrys1}d)  shows similar tendencies  as the regularity: it increases during loading and reduces during unloading. The~remaining temperature at the end of the unloading phase is due to the produced dissipation (Equation~\eqref{di7}) which contributes to Equation~\eqref{balance3} and leads to temperature~increase.

\begin{figure}[H]
  \centering
  \includegraphics[width=\textwidth]{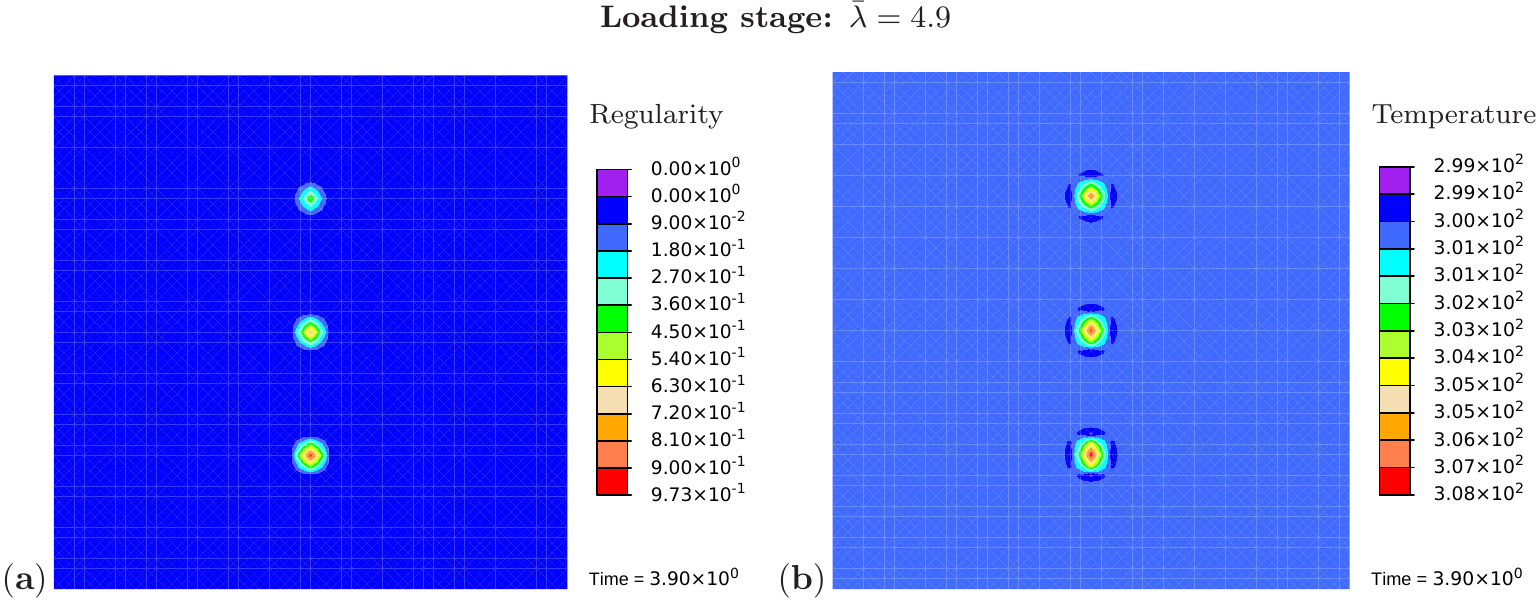}
  \caption{Simulation of the cyclic tensile test for a sample with three nuclei. Snapshots of the microstructure evolution during the loading stage ($\bar{\lambda} = 4.9$) showing (\textbf{a}) regularity distribution and (\textbf{b}) temperature distribution (in Kelvin).}
  \label{singlecrys21}
\end{figure}

\begin{figure}[H]
  \centering
  \includegraphics[width=\textwidth]{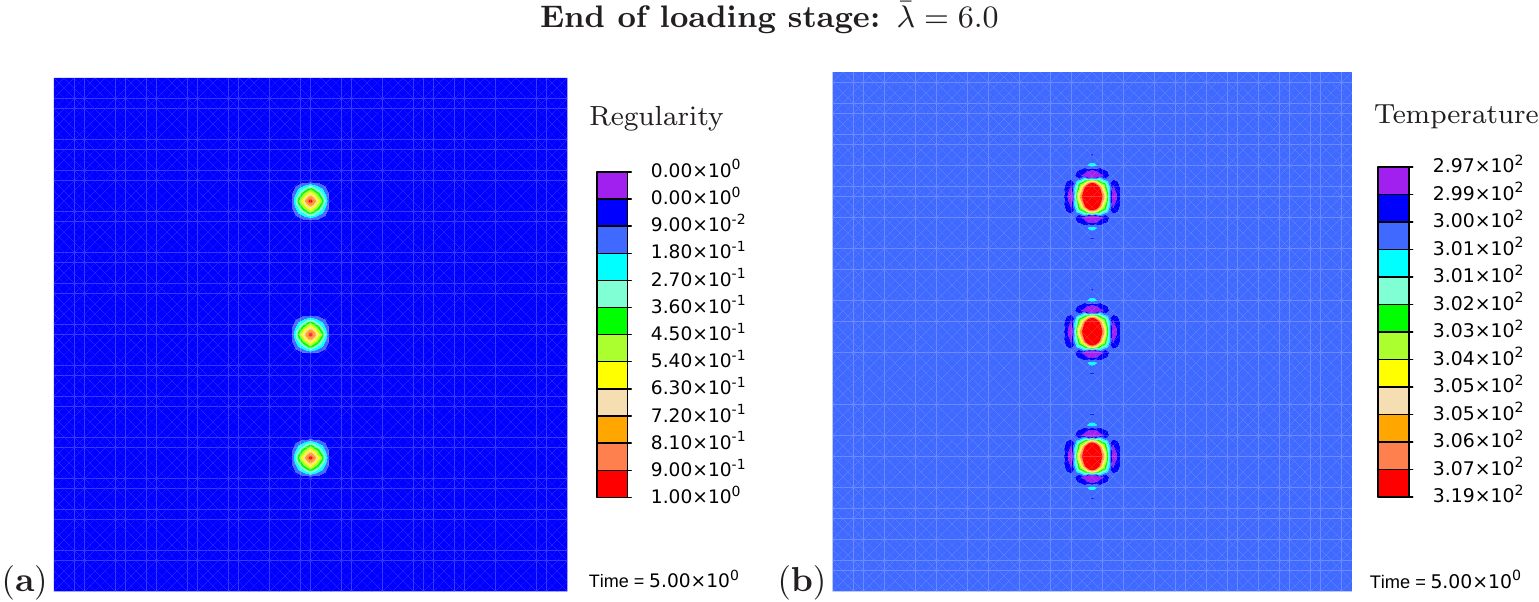}
  \caption{Simulation of the cyclic tensile test for a sample with three nuclei. Snapshots of the microstructure evolution at the end of the loading stage ($\bar{\lambda} = 6.0$) showing (\textbf{a}) regularity distribution and (\textbf{b}) temperature distribution (in Kelvin).}
  \label{singlecrys2}
\end{figure}

In addition, Figure~\ref{singlecrys21}a monitors the microstructure development during the loading phase at a stretch of $\bar{\lambda}=4.9$. The~resulting temperature distribution is shown in Figure~\ref{singlecrys21}b. The~crystalline regions build up and grow faster at the areas close to the element with a higher initial value of regularity. The~same holds for the temperature which rises more strongly in these areas.
Figure~\ref{singlecrys2} monitors the fully developed crystalline regions and the corresponding temperature distribution at the end of the loading phase ($\bar{\lambda} = 6.0$). The~highest temperature occurs directly in the crystallized region. Interestingly, the~area surrounding the crystal has cooled down. Equilibrium is established in the deformed state, with~the consequence that heat is absorbed from the surrounding~areas.

\unskip

\unskip

\subsection{Microstructure Evolution for a Sample with the Complex Initial~Microconfiguration}
\label{realmicro}
\begin{figure}[H]
  \centering
  \includegraphics[width=\textwidth]{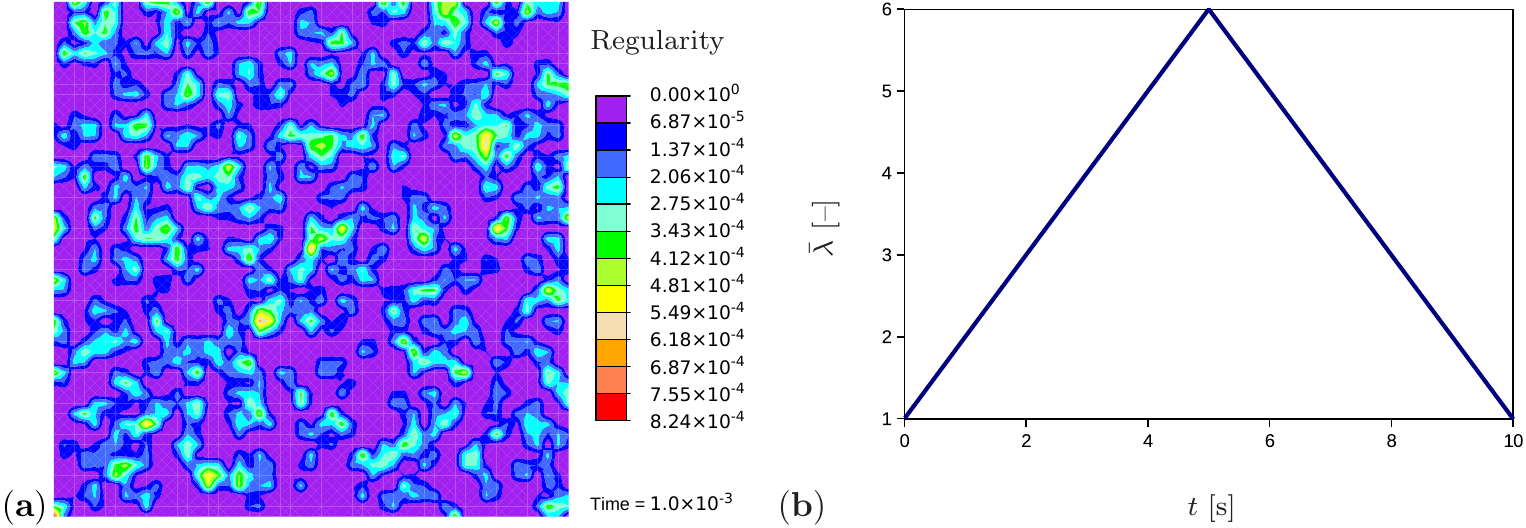}
  \caption{(\textbf{a}) Initial microstructure 
 with randomly distributed nuclei. (\textbf{b}) The applied stretch as a function of~time.}
  \label{rand1}
\end{figure}

A further example simulates the tensile test for a sample with randomly generated initial values of network regularity (Figure~\ref{rand1}a), which is the situation to be expected in a real polymer. The~initial values are generated in the range 
 [0, $1 \times 10^{-3}$] and randomly distributed over the sample as explained in Section~\ref{algor}. The~higher values of the regularity represent potential nuclei of crystalline regions. All remaining material and process parameters except the initial regularity are kept as in Section~\ref{1crys}. { Comparable simulations are performed using quadrilateral elements with the reduced integration; however, no evidence of volume locking is noticed.}

\begin{figure}[H]
  \centering
  \includegraphics[width=\textwidth]{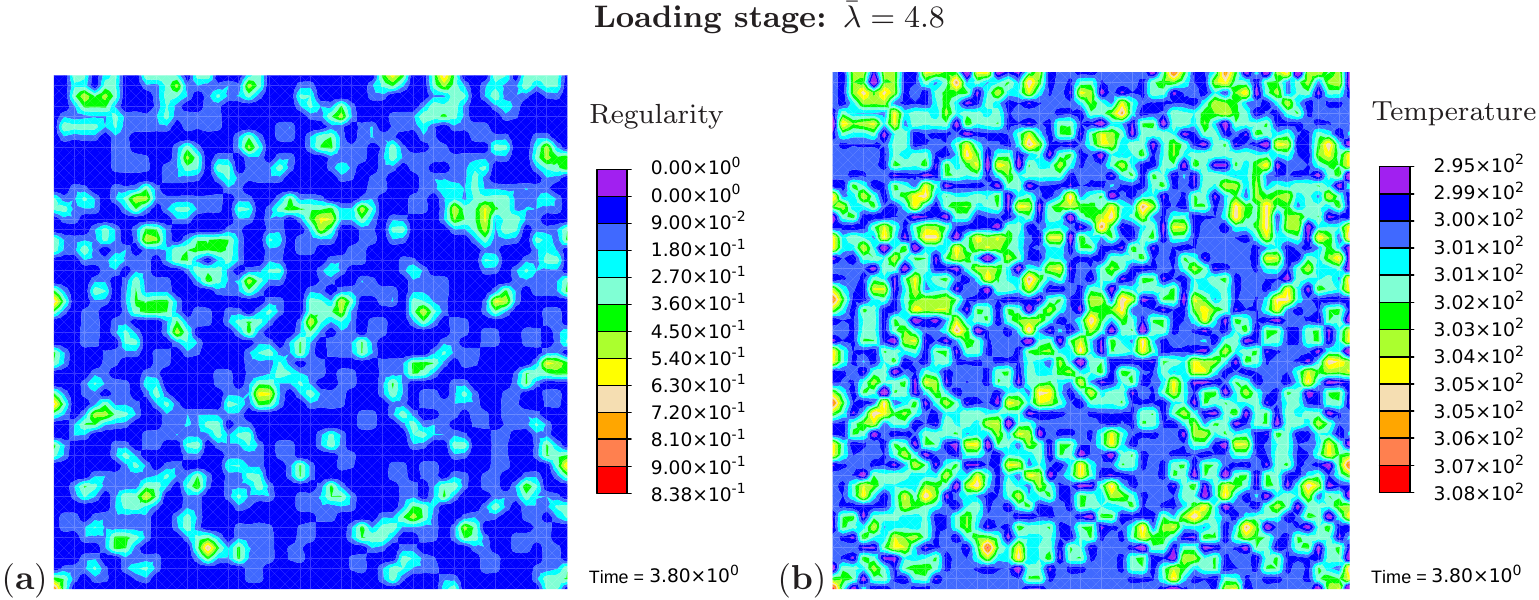}
  \caption{Cyclic tensile test with random initial microstructure. Snapshots of the microstructure evolution during the loading stage ($\bar{\lambda} = 4.8$) showing (\textbf{a}) regularity distribution and (\textbf{b}) temperature distribution (in Kelvin).}
  \label{rand2}
\end{figure}

\begin{figure}[H]
  \centering
  \includegraphics[width=\textwidth]{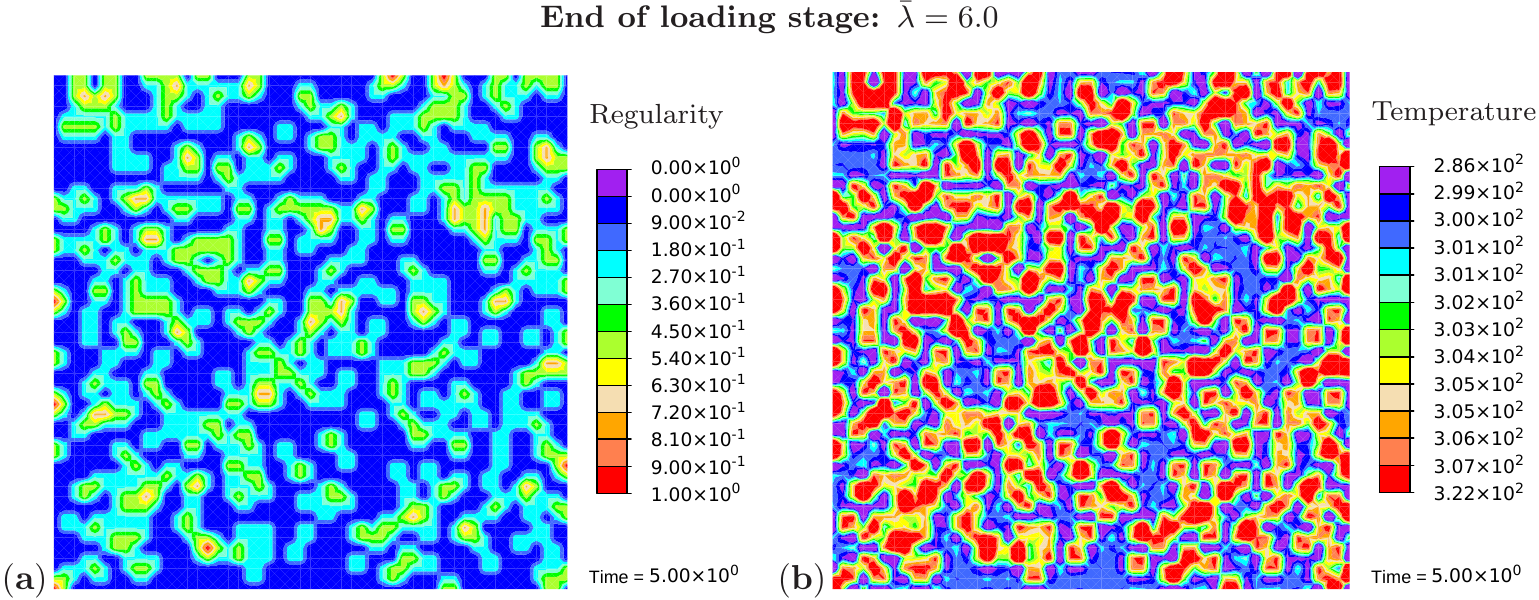}
  \caption{Cyclic tensile test with random initial microstructure. Snapshots of the microstructure evolution at the end of loading ($\bar{\lambda} = 6.0$) showing (\textbf{a}) regularity distribution and (\textbf{b}) temperature distribution (in Kelvin).}
  \label{rand3}
\end{figure}

\begin{figure}[H]
  \centering
  \includegraphics[width=\textwidth]{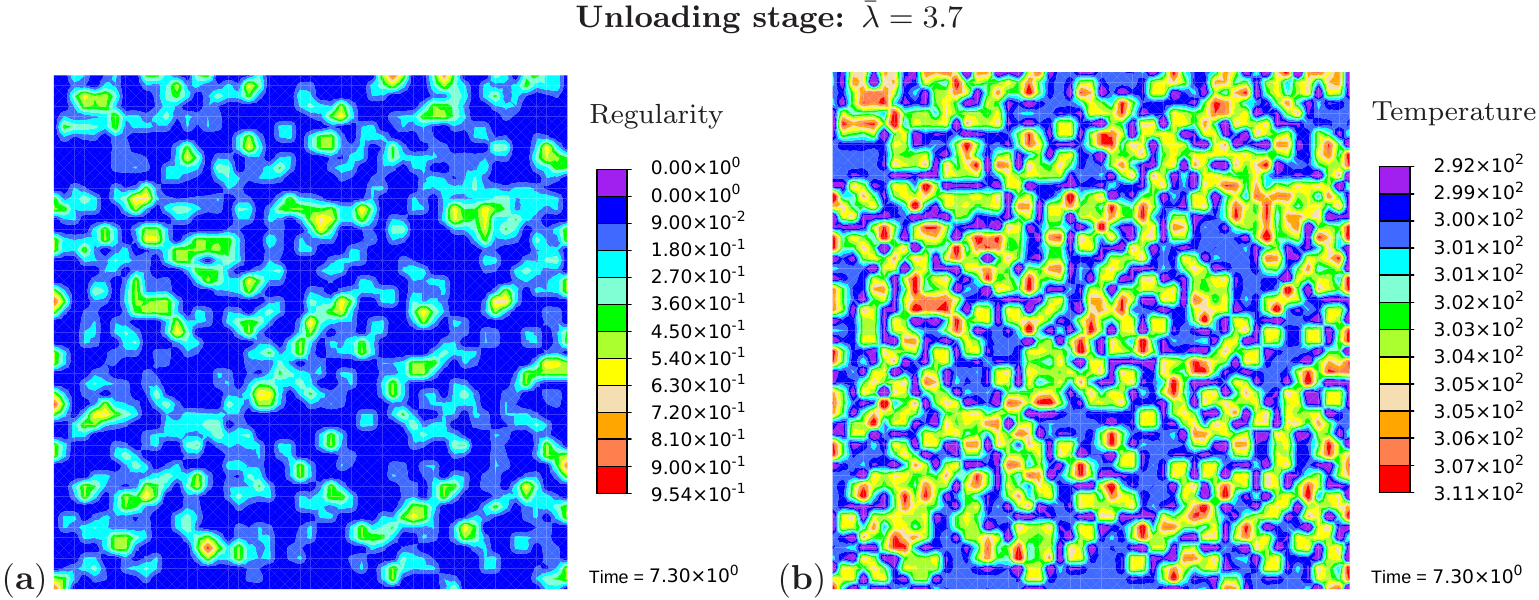}
  \caption{Cyclic tensile test with random initial microstructure. Snapshots of the microstructure evolution during the unloading stage ($\bar{\lambda} = 3.7$) showing (\textbf{a}) regularity distribution and (\textbf{b}) temperature distribution (in Kelvin).}
  \label{rand4}
\end{figure}

Three snapshots are chosen to display the microstructure development and the temperature distribution: Figure~\ref{rand2} shows the microstructure corresponding to an applied stretch of $\bar{\lambda} = 4.8$ during the loading phase, Figure~\ref{rand3} presents the situation at the end of the loading phase $\bar{\lambda} = 6.0$, and Figure~\ref{rand4} shows the microstructure for a stretch of $\bar{\lambda} = 3.7$ during unloading. The~color scale in Figure~\ref{rand1}a differs from the color scale in Figures~\ref{rand2}--\ref{rand4} in order to visualize the initial microstructure better. Crystalline regions (Figure~\ref{rand2}a) form during loading and this process generates heat so that an increase in temperature can be observed  (Figure~\ref{rand2}b). At~the highest stretch state, the~degree of crystallinity reaches about 15\% (Figure~\ref{rand3}a). The~highest temperatures are located in regions with a high density of crystals, since this is where the heat development of adjacent crystals accumulates. These~temperatures correspond to the red color in Figure~\ref{rand3}b. In~addition, it can be seen that the immediate surroundings of the crystals cool down (purple areas). As~already shown in the example with three crystals, a~withdrawal of heat from the amorphous regions also takes place during crystallization. Amorphous regions located at a larger distance from the crystals hardly experience any temperature changes and are shown in blue. During~the unloading phase the regression of the crystallization (Figure~\ref{rand4}a) absorbs heat, thereby reducing the temperature (Figure~\ref{rand4}b). The~comparison of the two states from Figures~\ref{rand2} and \ref{rand4} also demonstrates that the rate of the network regularity is higher during the loading stage than during the unloading stage at the same level of the stretch $\bar{\lambda}$.
\unskip

The simulations shown also enable an analysis of the effective material behavior, which is illustrated by examples of the crystallinity degree and of the effective temperature.
The crystallinity degree in the proposed model is defined as the volume fraction of the crystalline regions with the network regularity over 80$\%$. Its evolution compared to the experimental results is presented in Figure~\ref{average}a.
 The crystalline regions start to build at $\bar{\lambda} = 4.3$ and their volume fraction gradually increases up to the value of 15$\%$ at the end of the loading phase. The~crystallinity degree  decreases during the unloading stage and crystalline regions completely vanish at $\bar{\lambda} = 3.0$. The~rate of change during the loading phase is higher than during the unloading stage.  A~comparison shows that model for a chosen set of parameters predicts a slightly faster formation and degradation of the crystals 
than is observed in experiments. It also should be noted that data on the start of the crystallization, as~well as the value of the crystallinity degree at the peak are used to fit the parameters in Table~\ref{mat_par} and to fit the volume fraction of the sample with the initial regularity different than zero. 

{
The effective temperature $\Theta^{\mathrm{ef}}$ is  evaluated according to the principle of the volume averaging
\begin{equation}
  \Theta^{\mathrm{ef}} = \frac{1}{V} \int_{\mathcal{B}} \Theta \, \mathrm{d} \, V \text{ }
\end{equation}
and corresponding change with respect to the initial temperature is shown in Figure~\ref{average}b. The~temperature strongly increases after reaching $\bar{\lambda} = 4.3$ since the crystallization process starts. At~the end of the loading stage, the~model predicts a  temperature which  excellently agrees with the experimental value and corresponds to the temperature change of 6 K. During~the unloading phase, the~temperature gradually decreases with a moderate slope; however, it does not completely recover its initial value. This discrepancy can be explained by the fact that the present model simulates a closed system, whereas an exchange of the heat with the surroundings is possible in experiments~\cite{CANDAU2015244}.
\begin{figure}[H]
  \centering
  \includegraphics[width=\textwidth]{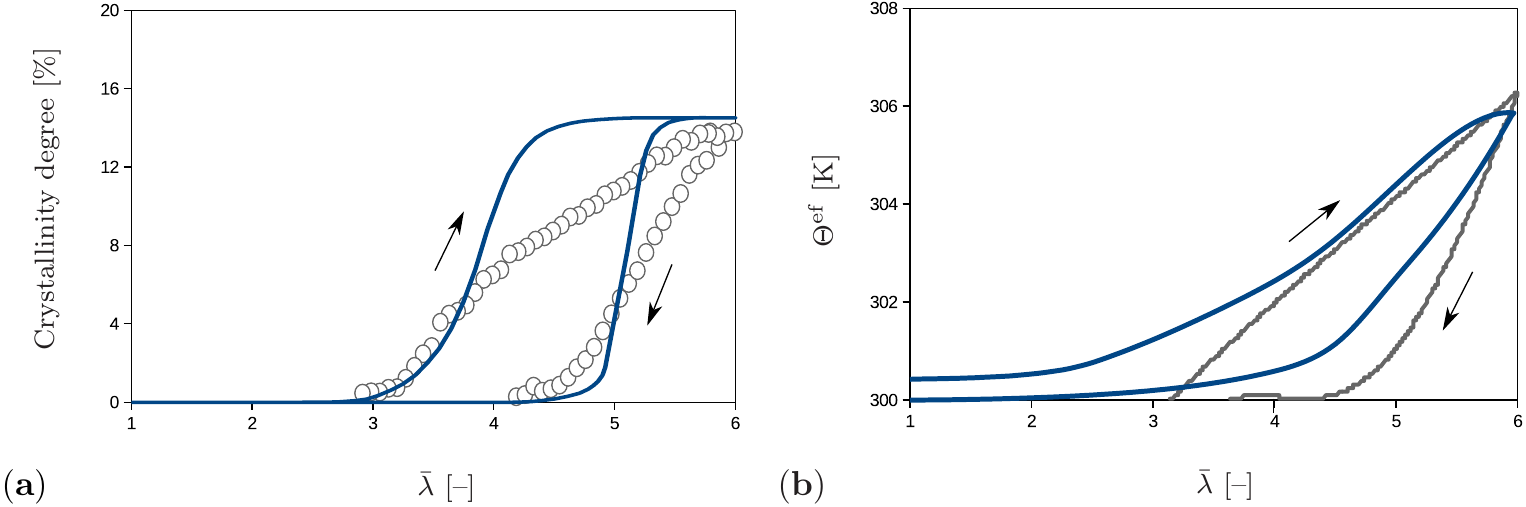}
  \caption{Comparison of the experimental results (in grey) and numerical results (in blue) for (\textbf{a}) the crystallinity degree and (\textbf{b}) the temperature change.}
  \label{average}
\end{figure}
}

\section{Conclusion and~Outlook}
This contribution presents a thermodynamically consistent model for the SIC in unfilled polymers under consideration of mechanical and thermal effects. The~kinematics is defined by distinguishing configurations due to elastic, crystalline, and thermal deformations, which yields the multiplicative split of the deformation gradient into three contributions. The~evolution equations of the internal variables describing the regularity of polymer network and the flexibility of polymer chains as a result of heat generation by crystallization are derived from the dissipation potential that has been established specifically for this type of material. The~assumption for Helmholtz energy includes the Arruda--Boyce elastic energy term, the~crystalline energy term, the~purely thermal energy part, and the mixed energy contribution dependent on temperature and thermal flexibility. Finally, the~weak form of the BVP including the balance of the linear momentum and balance of energy along with the suitable boundary conditions is defined for the purpose of the numerical implementation of the model proposed into the FEM-software. 

The application of the model is illustrated by examples dealing with the  microstructure evolution during a cyclic tensile test. In~the first example, the~sample contains three nuclei embedded in the amorphous matrix material. The~aim of this academic test is to investigate the material response affected by the SIC. The~numerical results show the evolution of the network regularity,  the~stress--stretch curve building a hysteresis, and the temperature change with similar trends as the regularity. In~addition, the~snapshot of the temperature distribution at the end of the loading path shows that  heat is generated in the crystalline region, whereas the adjacent amorphous area cools down. A~further example simulates the behavior of a sample with the configuration to be expected in a real amorphous polymer and enables the visualization of the SIC process by means of snapshots in several load stages. In~the temperature diagrams, the~heat generation and absorption due to the formation or degradation of the crystallization can be observed. It is noticeable that the temperature especially increases at places with clusters of crystalline regions, while the amorphous matrix outside the ``active'' areas hardly experiences any temperature changes at all. {The obtained results also serve as a basis for the evaluation of the crystallinity degree and of the effective temperature which plays an important role in the experimental validation.}

Apart from the issues mentioned, the~developed model also gives rise to other investigations. In~a first step, the~assumptions for the Helmholtz energy and the dissipation potential can be extended in order to simulate further effects occurring in filled and unfilled rubbers. Some important topics in this context would be the Mullins effect, a~deformation state beyond the elastic limit, induced~anisotropy, the~Gough--Joule effect, and heat conduction~\cite{juhre2013,voges2017}. In~addition, the~model proposed can be coupled to the phase--field approach in order to represent the biphasic nature of material better. Internal~variable $\chi$ would correspond to the order parameter in that case, and~its evolution could be controlled by the same dissipation potential as proposed in the present~work. 

\vspace{6pt} 

\paragraph{Funding}
This research was funded by the German Research Foundation (DFG) grant number KL 2678/5-1. The~same association along with TU Dortmund University has supported Open Access Publishing.

\paragraph{Acknowledgments}
We gratefully acknowledge the financial support by the German Research Foundation (DFG) and TU Dortmund University within the funding programme Open Access Publishing, as~well as the financial support by the DFG for the scientific project KL 2678/5-1. We furthermore thank J. Tiller and F. Katzenberg (TU~Dortmund University, Dortmund, Germany), and~S. Govindjee (University of California at Berkeley, Berkeley, USA) for their helpful discussions.


\bibliographystyle{acm}
\bibliography{literature}

\begin{thebibliography}{10}

\bibitem{ahzi2003}
{\sc Ahzi, S., Makradi, A., Gregory, R., and Edie, D.}
\newblock Modeling of deformation behavior and strain-induced crystallization
  in poly(ethylene terephthalate) above the glass transition temperature.
\newblock {\em Mech. Mater. 35}, 12 (2003), 1139--1148.

\bibitem{alfrey1942}
{\sc Alfrey, T., and Mark, H.}
\newblock A statistical treatment of crystallization phenomena in high
  polymers.
\newblock {\em J. Phys. Chem. 46}, 1 (1942), 112--118.

\bibitem{ARRUDA1993389}
{\sc Arruda, E.~M., and Boyce, M.~C.}
\newblock A three-dimensional constitutive model for the large stretch behavior
  of rubber elastic materials.
\newblock {\em J. Mech. Phys. Solids 41}, 2 (1993), 389--412.

\bibitem{AYGUN2020129}
{\sc Aygün, S., and Klinge, S.}
\newblock Continuum mechanical modeling of strain-induced crystallization in
  polymers.
\newblock {\em Int. J. Solids Struct. 196-197\/} (2020), 129--139.

\bibitem{BEHNKE201815}
{\sc Behnke, R., Berger, T., and Kaliske, M.}
\newblock Numerical modeling of time- and temperature-dependent strain-induced
  crystallization in rubber.
\newblock {\em Int. J. Solids Struct. 141-142\/} (2018), 15--34.

\bibitem{doi:10.5254/rct.13.86977}
{\sc Beurrot-Borgarino, S., Huneau, B., Verron, E., Thiaudière, D., Mocuta,
  C., and Zozulya, A.}
\newblock Characteristics of strain-induced crystallization in natural rubber
  during fatigue testing: In situ wide-angle x-ray diffraction measurements
  using synchrotron radiation.
\newblock {\em Rubber Chem. Technol. 87}, 1 (2014), 184--196.

\bibitem{BRADY2017181}
{\sc Brady, J., Dürig, T., Lee, P., and Li, J.-X.}
\newblock Chapter 7 - polymer properties and characterization.
\newblock In {\em Developing Solid Oral Dosage Forms (Second Edition)}, Y.~Qiu,
  Y.~Chen, G.~G. Zhang, L.~Yu, and R.~V. Mantri, Eds., second edition~ed.
  Academic Press, Boston, 2017, pp.~181--223.

\bibitem{bruening2012}
{\sc Brüning, K., Schneider, K., Roth, S., and Heinrich, G.}
\newblock Kinetics of strain-induced crystallization in natural rubber studied
  by {WAXD}: {D}ynamic and impact tensile experiments.
\newblock {\em Macromolecules 45\/} (10 2012), 7914--7919.

\bibitem{doi:10.1021/ma5006843}
{\sc Candau, N., Laghmach, R., Chazeau, L., Chenal, J.-M., Gauthier, C., Biben,
  T., and Munch, E.}
\newblock Strain-induced crystallization of natural rubber and cross-link
  densities heterogeneities.
\newblock {\em Macromolecules 47}, 16 (2014), 5815--5824.

\bibitem{CANDAU2015244}
{\sc Candau, N., Laghmach, R., Chazeau, L., Chenal, J.-M., Gauthier, C., Biben,
  T., and Munch, E.}
\newblock Influence of strain rate and temperature on the onset of strain
  induced crystallization in natural rubber.
\newblock {\em Eur. Polym. J. 64\/} (2015), 244--252.

\bibitem{doi:10.1098/rsta.2018.0067}
{\sc Carroll, M.~M.}
\newblock Molecular chain networks and strain energy functions in rubber
  elasticity.
\newblock {\em Philos. Trans. R. Soc. A 377}, 2144 (2019), 20180067.

\bibitem{Coleman1963}
{\sc Coleman, B.~D., and Noll, W.}
\newblock The thermodynamics of elastic materials with heat conduction and
  viscosity.
\newblock {\em Arch. Ration. Mech. An. 13}, 1 (Dec 1963), 167--178.

\bibitem{dargazany2014}
{\sc Dargazany, R., Khi\^em, V.~N., Poshtan, E.~A., and Itskov, M.}
\newblock Constitutive modeling of strain-induced crystallization in filled
  rubbers.
\newblock {\em Phys. Rev. E 89\/} (Feb 2014), 022604.

\bibitem{de2011computational}
{\sc de~Souza~Neto, E., Peric, D., and Owen, D.}
\newblock {\em Computational Methods for Plasticity: Theory and Applications}.
\newblock Wiley, 2011.

\bibitem{doll2000}
{\sc Doll, S., Hauptmann, R., Schweizerhof, K., and Freischläger, C.}
\newblock On volumetric locking of low order solid and solid-shell elements for
  finite elastoviscoplastic deformations and selective reduced integration.
\newblock {\em Eng. Comput. 17\/} (11 2000).

\bibitem{doufas1999}
{\sc Doufas, A.~K., Dairanieh, I.~S., and McHugh, A.~J.}
\newblock A continuum model for flow-induced crystallization of polymer melts.
\newblock {\em J. Rheol. 43}, 1 (1999), 85--109.

\bibitem{ELGUEDJ2014388}
{\sc Elguedj, T., and Hughes, T.}
\newblock Isogeometric analysis of nearly incompressible large strain
  plasticity.
\newblock {\em Comput. Methods Appl. Mech. Eng. 268\/} (2014), 388--416.

\bibitem{felder2019}
{\sc Felder, S., Holthusen, H., Hesseler, S., Pohlkemper, F., Simon, J.-W.,
  Gries, T., and Reese, S.}
\newblock A finite strain thermo-mechanically coupled material model for
  semi-crystalline polymers.
\newblock In {\em XV International Conference on Computational Plasticity :
  Fundamentals and Applications\/} (2019), E.~Onatea, D.~R.~J. Owen, D.~Peric,
  M.~Chiumenti, and E.~de~Souza~Neto, Eds., CIMNE, pp.~249--260.

\bibitem{doi:10.1063/1.1746537}
{\sc Flory, P.~J.}
\newblock Thermodynamics of crystallization in high polymers. {I}.
  {C}rystallization induced by stretching.
\newblock {\em J. Phys. Chem. 15}, 6 (1947), 397--408.

\bibitem{flory1949}
{\sc Flory, P.~J.}
\newblock Thermodynamics of crystallization in high polymers. {IV}. {A} theory
  of crystalline states and fusion in polymers, copolymers, and their mixtures
  with diluents.
\newblock {\em J. Phys. Chem. 17}, 3 (1949), 223--240.

\bibitem{TF9615700829}
{\sc Flory, P.~J.}
\newblock Thermodynamic relations for high elastic materials.
\newblock {\em Trans. Faraday Soc. 57\/} (1961), 829--838.

\bibitem{gasser2002}
{\sc Gasser, T., and Holzapfel, G.}
\newblock A rate-independent elastoplastic constitutive model for biological
  fiber-reinforced composites at finite strains: Continuum basis, algorithmic
  formulation and finite element implementation.
\newblock {\em Comput. Mech. 29\/} (10 2002), 340--360.

\bibitem{doi:10.1063/1.5088299}
{\sc Gschwandl, M., Kerschbaumer, R.~C., Schrittesser, B., Fuchs, P.~F.,
  Stieger, S., and Meinhart, L.}
\newblock Thermal conductivity measurement of industrial rubber compounds using
  laser flash analysis: Applicability, comparison and evaluation.
\newblock {\em AIP Conf. Proc. 2065}, 1 (2019), 030041.

\bibitem{hackl2008}
{\sc Hackl, K., and Fischer, F.~D.}
\newblock On the relation between the principle of maximum dissipation and
  inelastic evolution given by dissipation potentials.
\newblock {\em Philos. Trans. R. Soc. A 464}, 2089 (2008), 117--132.

\bibitem{Hajidehi2017}
{\sc Hajidehi, M.~R., and Stupkiewicz, S.}
\newblock Gradient-enhanced model and its micromorphic regularization for
  simulation of lüders-like bands in shape memory alloys.
\newblock {\em Int. J. Solids Struct. 135\/} (11 2017), 208--218.

\bibitem{Hartmann2012}
{\sc Hartmann, S.}
\newblock Comparison of the multiplicative decompositions $\bs{F} =
  \bs{F}_{\Theta} \bs{F}_{\mathrm{m}}$ and $\bs{F} = \bs{F}_{\mathrm{m}}
  \bs{F}_{\Theta}$ in finite strain thermo- elasticity.
\newblock Tech. rep., Faculty of Mathematics/Computer Science and Mechanical
  Engineering, Clausthal University of Technology, 01 2012.

\bibitem{HARTMANN20032767}
{\sc Hartmann, S., and Neff, P.}
\newblock Polyconvexity of generalized polynomial-type hyperelastic strain
  energy functions for near-incompressibility.
\newblock {\em Int. J. Solids Struct. 40}, 11 (2003), 2767--2791.

\bibitem{holzapfel1996}
{\sc Holzapfel, G.~A., and Simo, J.~C.}
\newblock Entropy elasticity of isotropic rubber-like solids at finite strains.
\newblock {\em Comput. Methods Appl. Mech. Eng. 132\/} (05 1996), 17--44.

\bibitem{horgan2010}
{\sc Horgan, C., and Murphyj, J.}
\newblock Simple shearing of incompressible and slightly compressible isotropic
  nonlinearly elastic materials.
\newblock {\em J. Elast. 98\/} (02 2010), 205--221.

\bibitem{hoss2010}
{\sc Hoss, L., and Marczak, R.}
\newblock A new constitutive model for rubber-like materials.
\newblock {\em AMCA 29\/} (01 2010), 2759--2773.

\bibitem{huneau2011}
{\sc Huneau, B.}
\newblock Strain-induced crystallization of natural rubber: a review of x-ray
  diffraction investigations.
\newblock {\em Rubber Chem. Technol. 84}, 3 (2011), 425--452.

\bibitem{doi:10.1177/1081286511429886}
{\sc Itskov, M., Dargazany, R., and Hörnes, K.}
\newblock Taylor expansion of the inverse function with application to the
  langevin function.
\newblock {\em Math. Mech. Solids 17}, 7 (2012), 693--701.

\bibitem{Jedynak}
{\sc Jedynak, R.}
\newblock A comprehensive study of the mathematical methods used to approximate
  the inverse langevin function.
\newblock {\em Math. Mech. Solids 24}, 7 (11 2018), 1992--2016.

\bibitem{juhre2013}
{\sc Juhre, D., Raghunath, R., Klueppel, M., and Lorenz, H.}
\newblock A microstructure-based model for filled elastomers including
  time-dependent effects.
\newblock {\em Constitutive Models for Rubber VIII\/} (04 2013), 293--298.

\bibitem{doi:10.1080/15376494.2020.1762952}
{\sc Kadapa, C., and Hossain, M.}
\newblock A linearized consistent mixed displacement-pressure formulation for
  hyperelasticity.
\newblock {\em Mech. Adv. Mater. Struc. 0}, 0 (2020), 1--18.

\bibitem{khiem2019}
{\sc Khi\^em, V.~N., Balandraud, X., and Itskov, M.}
\newblock Thermo-micromechanics of strain-induced crystallization.
\newblock In {\em Constitutive Models for Rubber XI\/} (London, 06 2019),
  B.~Huneau, J.-B. Le~Cam, Y.~Marco, and E.~Verron, Eds., Taylor \& Francis
  Ltd, pp.~30--35.

\bibitem{KHIEM2018350}
{\sc Khi\^em, V.~N., and Itskov, M.}
\newblock Analytical network-averaging of the tube model: Strain-induced
  crystallization in natural rubber.
\newblock {\em J. Mech. Phys. Solids 116\/} (2018), 350--369.

\bibitem{klinge2012}
{\sc Klinge, S., Bartels, A., and Steinmann, P.}
\newblock The multiscale approach to the curing of polymers incorporating
  viscous and shrinkage effects.
\newblock {\em Int. J. Solids Struct. 49}, 26 (2012), 3883--3900.

\bibitem{doi:10.1098/rspa.2014.0994}
{\sc Klinge, S., Hackl, K., and Renner, J.}
\newblock A mechanical model for dissolution--precipitation creep based on the
  minimum principle of the dissipation potential.
\newblock {\em Philos. Trans. R. Soc. A 471}, 2180 (2015), 20140994.

\bibitem{Kojio2011}
{\sc Kojio, K., Matsuo, K., Motokucho, S., Yoshinaga, K., Shimodaira, Y., and
  Kimura, K.}
\newblock Simultaneous small-angle {X}-ray scattering/wide-angle {X}-ray
  diffraction study of the microdomain structure of polyurethane elastomers
  during mechanical deformation.
\newblock {\em Polym. J. 43\/} (06 2011), 692--699.

\bibitem{kroon2010}
{\sc Kroon, M.}
\newblock A constitutive model for strain-crystallising rubber-like materials.
\newblock {\em Mech. Mater. 42}, 9 (2010), 873--885.

\bibitem{kurth2002continuum}
{\sc Kurth, J.~A., and Haupt, P.}
\newblock {\em Continuum Mechanics and Theory of Materials}.
\newblock Advanced Texts in Physics. Springer Berlin Heidelberg, 2002.

\bibitem{10.5254/1.3525684}
{\sc Le~Cam, J.-B.}
\newblock A review of volume changes in rubbers: the effect of stretching.
\newblock {\em Rubber Chem. Technol. 83}, 3 (09 2010), 247--269.

\bibitem{doi:10.1111/str.12256}
{\sc Le~Cam, J.-B.}
\newblock Strain-induced crystallization in rubber: A new measurement
  technique.
\newblock {\em Strain 54}, 1 (2018), e12256.

\bibitem{10.1007/978-3-319-95074-7_11}
{\sc Le~Cam, J.-B.}
\newblock Measuring strain-induced crystallinity in rubbers from {IR}
  thermography.
\newblock In {\em Residual Stress, Thermomechanics {\&} Infrared Imaging,
  Hybrid Techniques and Inverse Problems, Volume 7\/} (Cham, 2019), A.~Baldi,
  S.~Quinn, X.~Balandraud, J.~M. Dulieu-Barton, and S.~Bossuyt, Eds., Springer
  International Publishing, pp.~57--62.

\bibitem{LEGAC2018209}
{\sc {Le Gac}, P.-Y., Albouy, P.-A., and Petermann, D.}
\newblock Strain-induced crystallization in an unfilled polychloroprene rubber:
  Kinetics and mechanical cycling.
\newblock {\em Polymer 142\/} (2018), 209--217.

\bibitem{LIAO2020103263}
{\sc Liao, Z., Hossain, M., Yao, X., Mehnert, M., and Steinmann, P.}
\newblock On thermo-viscoelastic experimental characterization and numerical
  modelling of vhb polymer.
\newblock {\em Int. J. Nonlin. Mech. 118\/} (2020), 103263.

\bibitem{Maeda2015}
{\sc Maeda, N., Fujikawa, M., Makabe, C., Yamabe, J., Kodama, Y., and Koishi,
  M.}
\newblock {\em Performance evaluation of various hyperelastic constitutive
  models of rubbers}.
\newblock CRC Press, 07 2015, pp.~271--277.

\bibitem{MAHNKEN20132003}
{\sc Mahnken, R.}
\newblock Thermodynamic consistent modeling of polymer curing coupled to
  visco–elasticity at large strains.
\newblock {\em Int. J. Solids Struct. 50}, 13 (2013), 2003--2021.

\bibitem{MANSOURI20144316}
{\sc Mansouri, M., and Darijani, H.}
\newblock Constitutive modeling of isotropic hyperelastic materials in an
  exponential framework using a self-contained approach.
\newblock {\em Int. J. Solids Struct. 51}, 25 (2014), 4316--4326.

\bibitem{MCBRIDE20112116}
{\sc McBride, A., Javili, A., Steinmann, P., and Bargmann, S.}
\newblock Geometrically nonlinear continuum thermomechanics with surface
  energies coupled to diffusion.
\newblock {\em J. Mech. Phys. Solids 59}, 10 (2011), 2116 -- 2133.

\bibitem{mistry2014}
{\sc Mistry, S.~J., and Govindjee, S.}
\newblock A micro-mechanically based continuum model for strain-induced
  crystallization in natural rubber.
\newblock {\em Int. J. Solids Struct. 51}, 2 (2014), 530--539.

\bibitem{moler2003}
{\sc Moler, C., and Van~Loan, C.}
\newblock Nineteen dubious ways to compute the exponential of a matrix,
  twenty-five years later.
\newblock {\em SIAM Rev. 45}, 1 (2003), 3--49.

\bibitem{nateghi2018}
{\sc Nateghi, A., Dal, H., Keip, M.-A., and Miehe, C.}
\newblock An affine microsphere approach to modeling strain-induced
  crystallization in rubbery polymers.
\newblock {\em Contin. Mech. Thermodyn. 30}, 3 (May 2018), 485--507.

\bibitem{NEGAHBAN20002811}
{\sc Negahban, M.}
\newblock Modeling the thermomechanical effects of crystallization in natural
  rubber: {III}. {M}echanical properties.
\newblock {\em Int. J. Solids Struct. 37}, 20 (05 2000), 2811--2824.

\bibitem{Nie2015}
{\sc Nie, Y.}
\newblock Strain-induced crystallization of natural rubber/zinc dimethacrylate
  composites studied using synchrotron x-ray diffraction and molecular
  simulation.
\newblock {\em J. Polym. Res. 22\/} (02 2015).

\bibitem{doi:10.1098/rspa.1972.0096}
{\sc Ogden, R.~W., and Hill, R.}
\newblock Large deformation isotropic elasticity: on the correlation of theory
  and experiment for compressible rubberlike solids.
\newblock {\em Proc. R. Soc. A 328}, 1575 (1972), 567--583.

\bibitem{QU20095053}
{\sc Qu, L., Huang, G., Liu, Z., Zhang, P., Weng, G., and Nie, Y.}
\newblock Remarkable reinforcement of natural rubber by deformation-induced
  crystallization in the presence of organophilic montmorillonite.
\newblock {\em Acta Mater. 57}, 17 (2009), 5053--5060.

\bibitem{Raniecki1991}
{\sc Raniecki, B., and Bruhns, O.}
\newblock Thermodynamic reference model for elastic-plastic solids undergoing
  phase transformations.
\newblock {\em Arch. Mech. 43}, 2-3 (1991), 343--376.

\bibitem{RAO20011149}
{\sc Rao, I., and Rajagopal, K.}
\newblock A study of strain-induced crystallization of polymers.
\newblock {\em Int. J. Solids Struct. 38}, 6 (2001), 1149--1167.

\bibitem{REESE2003909}
{\sc Reese, S.}
\newblock A micromechanically motivated material model for the
  thermo-viscoelastic material behaviour of rubber-like polymers.
\newblock {\em Int. J. Plast. 19}, 7 (2003), 909--940.

\bibitem{reese_govindjee}
{\sc Reese, S., and Govindjee, S.}
\newblock Theoretical and numerical aspects in the thermo-viscoelastic material
  behaviour of rubber-like polymers.
\newblock {\em Mech. Time Depend. Mater. 1\/} (12 1997), 357--396.

\bibitem{Shahzad2015}
{\sc Shahzad, M., Kamran, A., Siddiqui, M.~Z., and Farhan, M.}
\newblock Mechanical characterization and {FE} modelling of a hyperelastic
  material.
\newblock {\em Mater. Res. 18\/} (10 2015), 918--924.

\bibitem{SIMO1991273}
{\sc Simo, J.~C., and Taylor, R.~L.}
\newblock Quasi-incompressible finite elasticity in principal stretches.
  continuum basis and numerical algorithms.
\newblock {\em Comput. Methods Appl. Mech. Eng. 85}, 3 (1991), 273--310.

\bibitem{sommer2016}
{\sc Sommer, S., Raidt, T., Fischer, B., Katzenberg, F., Tiller, J., and Koch,
  M.}
\newblock {TH}z-spectroscopy on high density polyethylene with different
  crystallinity.
\newblock {\em J. Infrared Millim. Terahertz Waves 37\/} (2016), 189–197.

\bibitem{SPRATTE201712}
{\sc Spratte, T., Plagge, J., Wunde, M., and Klüppel, M.}
\newblock Investigation of strain-induced crystallization of carbon black and
  silica filled natural rubber composites based on mechanical and temperature
  measurements.
\newblock {\em Polymer 115\/} (2017), 12--20.

\bibitem{Steinmann2012}
{\sc Steinmann, P., Hossain, M., and Possart, G.}
\newblock Hyperelastic models for rubber-like materials: Consistent tangent
  operators and suitability for treloar's data.
\newblock {\em Arch. Appl. Mech. 82\/} (09 2012).

\bibitem{toki2003}
{\sc Toki, S., Sics, I., Ran, S., Liu, L., and Hsiao, B.~S.}
\newblock Molecular orientation and structural development in vulcanized
  polyisoprene rubbers during uniaxial deformation by in situ synchrotron
  {X}-ray diffraction.
\newblock {\em Polymer 44}, 19 (2003), 6003--6011.

\bibitem{tosaka2004}
{\sc Tosaka, M., Murakami, S., Poompradub, S., Kohjiya, S., Ikeda, Y., Toki,
  S., Sics, I., and Hsiao, B.~S.}
\newblock Orientation and crystallization of natural rubber network as revealed
  by waxd using synchrotron radiation.
\newblock {\em Macromolecules 37}, 9 (2004), 3299--3309.

\bibitem{TOSAKA2012864}
{\sc Tosaka, M., Senoo, K., Sato, K., Noda, M., and Ohta, N.}
\newblock Detection of fast and slow crystallization processes in
  instantaneously-strained samples of cis-1,4-polyisoprene.
\newblock {\em Polymer 53}, 3 (2012), 864--872.

\bibitem{TF9545000881}
{\sc Treloar, L. R.~G.}
\newblock The photoelastic properties of short-chain molecular networks.
\newblock {\em Trans. Faraday Soc. 50\/} (1954), 881--896.

\bibitem{voges2017}
{\sc Voges, J., Makvandi, R., and Juhre, D.}
\newblock Numerical investigation of the phase evolution in polymer blends
  under external mechanical loadings.
\newblock {\em Tech. Mech. 37\/} (01 2017), 37--47.

\bibitem{Yamamoto2019}
{\sc Yamamoto, T.}
\newblock Molecular dynamics simulation of stretch-induced crystallization in
  polyethylene: Emergence of fiber structure and molecular network.
\newblock {\em Macromolecules 52}, 4 (2019), 1695--1706.

\bibitem{doi:10.1063/1.5063384}
{\sc Zhang, Q., Li, X., and Yang, Q.}
\newblock Extracting the isotropic uniaxial stress-strain relationship of
  hyperelastic soft materials based on new nonlinear indentation strain and
  stress measure.
\newblock {\em AIP Advances 8}, 11 (2018), 115013.

\end{thebibliography}

\end{document}